\documentclass[twocolumn,showpacs,preprintnumbers,amsmath,amssymb,showkeys]{revtex4}
\usepackage{graphicx}
\usepackage{dcolumn}
\usepackage{bm}

\begin{document}
\title{Electronic Raman scattering in a multiband model for cuprate
superconductors}
\author{Ivan Kup\v{c}i\'{c}
and  Slaven Bari\v{s}i\'{c}} 
\affiliation{ Department of Physics, 
        Faculty of Science, POB 331,  HR-10\,002 Zagreb, Croatia}

\begin{abstract} 
Charge-charge, current-current and Raman correlation functions 
are derived  in a consistent way using the unified response theory.
 	The theory is based on the improved  description
of the conduction electron coupling to the external 
electromagnetic fields, distinguishing further the direct and 
indirect (assisted) scattering on the quasi-static disorder.
 	The two scattering channels are  distinguished in terms 
of the energy and momentum conservation laws.
	The theory is illustrated on the  Emery three-band model
for the normal state of the underdoped high-$T_c$ cuprates 
which includes the  incoherent electron scattering 
on the  disorder associated  with the quasi-static fluctuations
around the static antiferromagnetic (AF) ordering.  
	It is shown, for the first time consistently, 
that the incoherent indirect 
processes dominate the low-frequency part of the Raman spectra, 
while the long-range screening which is dynamic removes the 
long-range forces in the $A_{1g}$ channel. 
	In the mid-infrared
frequency range the coherent AF processes are dominant.
 	In contrast to the nonresonant $B_{1g}$ response, 
which is large  by itself, the resonant interband transitions  
enhance both the $A_{1g}$ and $B_{1g}$ Raman spectra
to comparable values, 
in good agreement with experimental observation.
	It is further argued that the AF correlations 
give rise to the mid-infrared peak in the $B_{1g}$ Raman spectrum, 
accompanied by a similar peak in the optical conductivity.
	The doping behavior of these peaks is shown to be correlated
with the linear doping dependence of the Hall number, as observed in all
underdoped high-$T_c$ compounds. 
\end{abstract}


\maketitle

\section{Introduction }
%
%

Multiband models often present several energy scales of the same order of
magnitude, related to various anticrossings of the bands.
	One such interesting example is the Emery model 
for the high-$T_c$ cuprates.
	The effective band structure of this model exhibits
hybridization gaps related to the anticrossings of three bands associated
with the CuO$_2$ unit cell, as well as the dimerization pseudogaps related 
to the antiferromagnetic (AF) fluctuations, all of the order of 0.1 eV.
	The obvious prerequisite for the understanding of the high-$T_c$ 
superconductivity, which in turn is associated with energies of the 
order of 0.01 eV, is the correct identification of the origin of the
0.1 eV energy scales. 
	In the attempt to distinguish among the 0.1 eV energy scales, 
one is left only with the difference in the associated behaviors in the 
momentum space, i.e. with the corresponding coherence factors,
to use the band language.
	As is well known, the coherence factors reflect the crystal symmetry
and experimental probes sensitive to the associated selection
rules, such as infrared conductivity and Raman scattering
\cite{Uchida,Cooper,Cardona,Reznik,Sugai,Sugai2,Opel,Hackl,Blumberg}, 
are well suited for the study of the coherence factors.
	The motivation of the present paper is to discuss theoretically
the existing Raman data from such a point of view.
	This is accompanied here by the solution of several long-standing
problems which concern the electronic Raman scattering in general.

More specifically, the experimental Raman investigations of the effects of
superconductivity on the Drude part of the $B_{2g}$  spectra of  
YBa$_2$Cu$_3$O$_{7-x}$ \cite{Opel} 
and Bi$_2$Sr$_2$Ca$_1$Cu$_{2}$O$_{8+x}$ \cite{Devereaux3}
confirmed the conclusions of other experiments 
\cite{Ding,Loram}
that the superconducting gap/pseudogap is of the order of 25 meV,
with a predominant  $d_{x^2-y^2}$ symmetry.
	In addition,  the $B_{1g}$ spectra in underdoped 
La$_{2-x}$Sr$_x$CuO$_4$ \cite{Naeini} and 
Bi$_2$Sr$_2$Ca$_1$Cu$_{2}$O$_{8+x}$ \cite{Sugai2}
compounds show at temperatures up to room temperature
a strong two-magnon peak at  0.1--0.3  eV and
a secondary structure at a frequency about three times lower.
	Both scales exhibit the same doping behavior.
	The  smaller scale is therefore usually associated with the 
single-paramagnon  AF pseudogap
\cite{Sugai,Sugai2,Loram,Naeini}.
	Similar scales appear in other experiments,
in particular in  measurements of the specific heat \cite{Loram}.
 	Equally important are the overdoped cuprates
where 0.1 eV energy scales are observed in featureless mid-infrared 
spectra in  optical conductivity and Raman
experiments \cite{Uchida,Cooper,Sugai,Sugai2}.
	The latter are usually associated with  the strong 
quasi-particle damping effects, that is, 
with the  scattering from the uncorrelated spin disorder, 
rather  than with the  AF paramagnons and the concomitant disorder.

The  small energy scales of the order of 0.1 eV and less occur in the
Emery three-band model for the high-$T_c$ cuprates in the limit
of large repulsive interaction on the Cu site 
\cite{Emery}.
	This interaction is renormalized out by introducing the 
auxiliary bosons \cite{Kotliar}, which forbid the double occupancy 
of the Cu site, i.e. by introducing the Mott charge correlations.
	The result for finite doping is the effective band structure
with bands broadened by the scattering of fermions on bosons.
	The single-particle dispersions obtained on the hole-doped side 
within the paramagnetic non-crossing-approximation (NCA) \cite{Niksic} 
or dynamical mean-field theory \cite{Zolfl} approaches are similar 
to those found by the simple mean-field slave-boson (MFSB) theory 
\cite{Kotliar,IMR}, when the latter is supplemented by harmonic boson
fluctuations around the mean-field saddle-point.
	The band dispersions introduce the non-magnetic energy scales of the
order of 0.1 eV and less, in particular through the splitting between
the resonant band and the main band.
	The band broadening $\Gamma  ({\bf k}, \omega)$ of 
the non-Fermi-liquid type is related to the inelastic scattering on anharmonic
(slave) bosons, which describe the Cu-O charge fluctuations irrespective of 
the spin.
	$\Gamma  ({\bf k}, \omega)$ is itself characterized by the energy
scales of 0.1 eV.
	The  Raman background corresponding to the charge fluctuations 
was evaluated within the NCA \cite{Niksic}.
	It reflects the same  non-magnetic 0.1 eV energy scales, in particular
through the processes of charge excitations from the main oxygen band to the 
resonant band.
	The agreement between the calculated single-particle \cite{IMR} 
and electron-hole Raman \cite{Niksic} properties 
and the corresponding ARPES \cite{Ino,Yoshida,Norman,Shabel} and 
Raman \cite{Cooper,Reznik} measurements on La$_{2-x}$Sr$_x$CuO$_4$
family of materials is  remarkable.

In this kind of approach the magnetic effects manifest as perturbations 
in terms of AF paramagnons \cite{DKS}.	
	The associated pseudogap energy $\Delta_{\rm AF}$
is well below 0.1 eV.
	Until now, the bosonic effects of paramagnons were estimated
only by omitting the band broadening due to bosonic charge fluctuations.
	This amounts to the use of the MFSB theory, supplemented by the 
coupling of the Fermi liquid to the paramagnons \cite{DKS}.
	Such an approximation conserves the 0.1 eV energy scales in the 
band dispersion and allows for the (in)elastic scattering on paramagnons.
	The corresponding inelastic processes turn out to be more
important \cite{DKS} on the hole-doped side than on the electron-doped 
side of the ``non-magnetic normal state''
extrapolated close below the superconducting $T_c$.
	The whole hierarchy of energy scales, and especially the assertion
that the relevant non-magnetic energy  scales are larger than $\Delta_{\rm AF}$,
which itself is larger than $T_c$, is obviously of essential importance 
for the understanding of high-$T_c$ superconductivity.

In order to investigate carefully the energy scale hierarchy, this paper
is focussed on the effect of the AF paramagnons on the Raman response,
introducing further simplifications which nevertheless conserve the main 
non-magnetic and magnetic scales at and below 0.1 eV.
	The  nonmagnetic scales below 0.1 eV are retained 
in the fermion dispersion.
	The AF correlations are described by the AF gap $\Delta_{\rm AF}$
instead of the pseudogap \cite{Friedel} and by the bosonic fluctuations 
(magnons) around the AF state.
	Both steps are usually considered as legitimate for temperatures
below $\Delta_{\rm AF}$ \cite{Brazovskii}.
	In this way, the inelastic scattering on magnons is neglected 
(in addition to that on charge fluctuations).
	This omits in particular the antiadiabatic magnon effects on 
the single-particle spectrum of holes \cite{DKS}
at energies very much below $\Delta_{\rm AF}$.
	The whole approach reduces in this way to the MFSB three-band theory 
with the AF dimerization which includes (only) the elastic scattering on the
(intrinsic AF and extrinsic) disorder.
	Even with such drastic simplifications the  problem is a serious one.

This article investigates in detail the  Raman spectra of the underdoped 
cuprates and  distinguishes among the coherence factors associated 
in the reciprocal space with the
non-magnetic and magnetic scales which appear in the problem.
	The usual Raman analysis of the high-$T_c$ cuprates starts 
from the simple  Abrikosov and Genkin approach \cite{Abrikosov}.
	The latter treats the  bi-linear Raman
excitations as non-resonant and calculates the Raman intraband 
contributions starting from the free electron limit
\cite{Kosztin,Zawadowski,Monien,Ruvalds,Devereaux1,Devereaux4,Devereaux2,Devereaux5,Niksic,Einzel}.
	This is replaced here by the description of the electron-photon
coupling effects which is more appropriate for the analysis of the relevant
coherence factors for a nearly half-filled tight-binding band.
 	In such a discussion it is obviously important  to 
account also  for the decoherence effects, associated at least with 
the elastic scattering of charge carriers on the quasi-static disorder.

Associated is the problem of the screening of the long-range Coulomb
forces in the presence of the disorder
\cite{Zawadowski,Monien,Ruvalds,Devereaux1,Devereaux4,Devereaux2,Devereaux5,Niksic,Einzel,Klein,Sherman2,BarisicIJMP,BarisicSSC}.
	This problem is usually treated in the Raman (and infrared)
analysis by the field-theory approximation (FTA). 
	In this approach the long-range  forces are screened
off by the coherent long-range screening
and the elastic scattering on the disorder is taken to 
break the translational symmetry, i.e. the momentum
conservation laws.
	The two steps may thus seem to be either contradictory or to 
amount to double counting. 
	By distinguishing the (direct) processes
with the quasi-particle momentum conservation, from the (indirect) processes,
which do not conserve the momentum, we show therefore that 
the two steps in question can be reconciled.
	The momentum conservation processes are subject to the 
coherent long-range screening, 
while the other processes do not imply long-range forces at all.

Being interested here primarily in the interband scales 
we extend the above single-band considerations to the multiband case.
	The role of interband transitions is twofold here.	
	First, the quasi-particles can be excited resonantly 
from the conduction band to the other bands.
	Second, the  excited quasi-particles relax back 
into the conduction band, assisted by
the elastic scattering  on the disorder.
	The former effect is treated by replacing  the usual 
static-Raman-vertex approximation (SRVA) by the 
elastic-Raman-vertex approximation (ERVA).
	This represents a natural extension of the recent multiband optical 
conductivity analysis \cite{KupcicPC,KupcicFA}
to the Raman case.
	Such  an approach gives access to the  most important non-magnetic
single-particle scales of the Emery model.
	On the other hand, it is shown that 
the additional elastic scattering on the disorder,
associated with the interband transitions, can be included into the 
(indirect) processes, which do not conserve the quasi-particle momentum.

The result of these steps is the theory of the electronic Raman scattering
in multiband models, the Emery model for the high-$T_c$ cuprates in particular,
which can be compared to the experimental findings.
	As the analogous theory applies also to the conductivity, this approach,
as a whole, establishes the relation among a number of measurable quantities
including the DC conductivity and the Hall number, all sensitive to the 
anomalous features in the quasi-particle spectrum close to the Fermi level, such as
hybridization/dimerization (pseudo)gaps and the van Hove singularities.
	It appears that the AF dimerization gap produces the intensity 
maximum in the 
$B_{1g}$ Raman channel as well as in the optical conductivity, while the 
low-lying $B_{2g}$ spectrum remains unaffected.
	In addition, the number of the van Hove singularities is doubled, which restores
approximately the local electron-hole symmetry in the conduction band.
	This agrees fully with the measured doping dependence of the Hall 
number in the underdoped electron- and hole-doped regimes
\cite{Uchida,Uchida2}.
	The small 0.1 eV energy scale observed in all these experiments in
the underdoped cuprates is thus associated  here with the AF dimerization 
rather than with the non-magnetic scales of the same order of magnitude.	
	Such interpretation requires however further confirmation
through the theory beyond the MFSB level.

The paper is organized as follows. 
	In Sec.~II the response of the electronic system to external 
transverse vector fields is formulated for a multiband
model and applied to the Emery three-band model
where the local field corrections are absent.
	The  contributions of the direct and indirect electron-hole pair 
excitations to the Raman correlation functions  are determined, 
including the screening by the multiband RPA (random phase approximation) 
dielectric function.
	The structure of the low-frequency (Drude) contribution to 
the Raman correlation functions is given in Secs.~III and IV.
	The relation between the ERVA and SRVA is  discussed in Sec.~V.
	The predictions of the model with AF correlations for the 
Hall number, the optical conductivity, and the corresponding  
contributions to the $B_{1g}$ and $B_{2g}$ Raman spectra 
are given in Sec.~VI, and compared to the experimental data.
 	Sec.~VII contains the concluding remarks.

\section{Multiband model Hamiltonian}
%
%
\subsection{Emery three-band model}
%
%

We consider the conduction electrons described by the 
reduced version of the quasi-two-dimensional Emery three-band model
\cite{Emery}, 
in which  the second-neighbor bond energy $t_{\rm pp}$ is set to zero, 
and  the short-range interactions $V_{\rm pd}$ and $V_{\rm pp}$  
are approximately included in the copper and oxygen 
single-particle  energies.
	The Hamiltonian is 
\begin{eqnarray}
H &=& { H}_0  +  H'_{1}  + H'_{2} + H^{\rm ext}.
\label{eq1}
 \end{eqnarray} 
	${ H}_0$ is the effective single-particle term. 
	The electron quasi-elastic scattering on the disorder is described 
by $H'_1$.	
	$H'_{2} = H_{\rm c}+ H_{\rm AF}$ represents 
the two-particle interactions, including both
the long-range Coulomb forces ($H_{\rm c}$) and the residual interactions 
responsible for the  AF correlations ($H_{\rm AF}$).
	$H^{\rm ext}$ describes the coupling of the conduction electrons
to the external fields.

Using the slave-boson approach 
to treat the limit of large Hubbard interaction
on the copper site $U_{\rm d}$, the effective 
MFSB single-particle Hamiltonian \cite{Kotliar}
can be written in the representation of the 
non-diagonal translationally invariant states as
\begin{eqnarray}
H_0 &=& 
\sum_{ll'{\bf k} \sigma} \big[ H_0^{ll'} ({\bf k}) 
l_{{\bf k} \sigma}^{\dagger} l'_{{\bf k} \sigma} + {\rm H.c.} \big],
\label{eq2}
\end{eqnarray}
%
%
with the orbital index $l, l' = d, \, p_x, \, p_y$.
	Here the diagonal and off-diagonal matrix elements have the well-known
form: $H_0^{ll} ({\bf k}) = E_l -2t_{\perp} \cos k_z a_3$,
$H_0^{{dp}_{\alpha}} ({\bf k}) =$  
\hbox{$2 {\rm i}t_{\rm pd}^{\rm eff} 
\sin \frac{1}{2} {\bf k} \cdot {\bf a}_{\alpha}$,} 
with $\alpha = x,y$, and
$H_0^{{p}_x {p}_y} ({\bf k}) =  -4t_{\rm pp} 
\sin \frac{1}{2} {\bf k} \cdot {\bf a}_{1}
\sin \frac{1}{2} {\bf k} \cdot {\bf a}_{2}$
(${\bf a}_1$, ${\bf a}_2$, and ${\bf a}_3$ are the primitive vectors of the
tetragonal lattice in question).
	$E_l$ are the renormalized site energies, 
$t_{\rm pd}^{\rm eff}$ is the renormalized first-neighbor bond-energy, 
$t_{\rm pp}$ is the second-neighbor bond-energy, and
$t_{\perp}$ is the interplane bond-energy.
	Using the transformations
\begin{eqnarray}
l_{{\bf k} \sigma}^{\dagger} &=& 
\sum_L U_{\bf k} (l,L)  L_{{\bf k} \sigma}^{\dagger}, 
\label{eq3}
 \end{eqnarray}
%
%
	$H_0$ is diagonalized in terms of three bands 
\begin{eqnarray}
H_0 &=& \sum_{L{\bf k} \sigma}   
E_L ({\bf k}) L_{{\bf k}  \sigma}^{\dagger} L_{{\bf k} \sigma},
\label{eq4}
\end{eqnarray}
%
%
with the band indices $L = c$ for the nearly half filled (conduction) 
bonding band and $L = N, P$ for the non-bonding and antibonding bands
(which are empty in the hole picture used here).
	For $t_{\rm pp}=0$, the structure of $E_L ({\bf k}) $ 
and $U_{\bf k} (l,L)$ is well known \cite{Kotliar,KupcicPRB2}.

The effects of the AF correlations on the Raman spectral functions is
approximated here by replacing the coupling of the conduction 
band electrons to the AF fluctuations by their coupling to the 
${\bf Q}_{\rm AF}$ mode, which is  taken as frozen in.
	The effect of bosons  with the wave vectors close to 
${\bf Q}_{\rm AF}$ on the quasi-particle dispersion is thus neglected, 
i.e. the pseudogap is replaced by the gap $\Delta ({\bf k})$ involved 
in $H_{\rm AF}$ \cite{Friedel,DKS},
\begin{eqnarray}
H_{\rm AF} &=& \sum_{{\bf k} \sigma }  \big[ \Delta ({\bf k})
c^{\dagger}_{{\bf k}  \sigma} 
c_{{\bf k} \pm {\bf Q}_{\rm AF} \underline{\sigma}} + {\rm H. c.} \big].
\label{eq5}
\end{eqnarray} 	
%
%
	On the other hand, the life-time effects associated with slow AF 
fluctuations can be included in the $H'_1 $ quasi-elastic scattering on the
disorder \cite{Mahan,Abrikosov2},
\begin{eqnarray}
H'_1  &=& 
\sum_{ L{\bf k} {\bf k}'  \sigma } 
V_1^{LL} ({\bf k}-{\bf k}') 
L_{{\bf k} \sigma}^{\dagger} L_{ {\bf k}' \sigma}.
\label{eq8}
\end{eqnarray}
%
%
	This implies the adiabatic approximation in the quasi-particle 
scattering on bosons, i.e. the boson frequency lower than the temperature
of interest \cite{DKS}.
 	As already  pointed out in Introduction, 
the corresponding corrections are not expected to  
affect much the conclusions which concern the 0.1 eV scale in the underdoped
compounds, below the two-magnon resonance \cite{Sugai,Sugai2}.
	This is the range to which we restrict ourselves here,
while discussing some basic questions, which concern the 
Raman scattering itself.

Finally, the long-range forces are given by 
\begin{eqnarray}
 H_{\rm c} &=& 
\sum_{  {\bf q} \neq 0}  \frac{2 \pi}{vq^2}  
\hat{q} (-{\bf q})  \hat{q} ({\bf q}),
\label{eq6}
\end{eqnarray}
%
%
with $\hat{q}({\bf q}) $ being the charge density operator, 
\begin{eqnarray}
\hat{q} ({\bf q})  &=&  \sum_{LL'} \sum_{{\bf k} \sigma}
eq^{LL'} ({\bf k},{\bf k}+{\bf q})  L_{{\bf k} \sigma}^{\dagger} 
L'_{{\bf k}+{\bf q}  \sigma}, 
\label{eq7} 
\end{eqnarray}
%
%
and the $q^{LL'} ({\bf k},{\bf k}+{\bf q})$ are the related dimensionless
intra- and interband charge vertices [see Appendix C and Eq.~(\ref{eq12})].

\subsection{Electromagnetic coupling}
%
%

The coupling of the conduction electrons to the  electromagnetic  fields polarized 
in the $\alpha$ and/or $\beta$ direction 
follows from the minimal gauge-invariant substitution
\cite{Pines,KupcicPB,KupcicPC} 
\begin{eqnarray}
 && H^{\rm ext}  =   H^{\rm ext}_1 +  H^{\rm ext}_2 =
-\frac{1}{c} \sum_{{\bf q} \alpha} 
A_{\alpha} ({\bf q}) \hat{J}_{\alpha} (-{\bf q}) 
 \nonumber  \\ 
&& \hspace{5mm}
- \frac{e^2}{2mc^2} \sum_{{\bf q} {\bf q}'\alpha \beta} 
A_{\alpha} ({\bf q}-{\bf q}')
A_{\beta} ({\bf q}') 
\hat{\gamma}_{\alpha \beta} ( -{\bf q};2).   
\label{eq9} 
 \end{eqnarray} 
%
%
	Here 
\begin{eqnarray}
\hat{J}_{\alpha} ({\bf q})  &=&  \sum_{LL'} \sum_{{\bf k} \sigma}
J^{LL'}_{\alpha} ({\bf k})  L_{{\bf k} \sigma}^{\dagger} 
L'_{{\bf k}+{\bf q}  \sigma}, 
 \nonumber  \\
\hat{\gamma}_{\alpha \beta} ({\bf q};2)  &=&  
\sum_{LL'} \sum_{{\bf k} \sigma}
\gamma^{LL'}_{\alpha \beta } ({\bf k};2)  L_{{\bf k} \sigma}^{\dagger} 
L'_{{\bf k}+{\bf q}   \sigma}, 
\label{eq10} 
\end{eqnarray}
%
%
are, respectively, the current density and bare
Raman  density operators \cite{Abrikosov,KupcicPC}.	
	The explicit form of the current vertices,  
$J^{LL'}_{\alpha} ({\bf k})$, and the bare Raman vertices, 
$\gamma^{LL'}_{\alpha \beta} ({\bf k};2)$ 
for the $t_{\rm pp}= 0$ Emery three-band model are given in  Appendix A.

The coupling (\ref{eq9}) can be completed with the coupling to the external 
scalar fields $V^{\rm ext} ({\bf q})$,
\begin{eqnarray}
H^{\rm ext}_0  &=& 
\sum_{{\bf q} } V^{\rm ext} ({\bf q}) \hat{q} (-{\bf q}),      
\label{eq11} 
\end{eqnarray} 
%
%
used in the longitudinal response theory (see Appendix C).
	It is important to notice that, due to the absence of 
the local field corrections \cite{Adler,ZBB}
in the Emery model, the long-wavelength charge vertices 
(${\bf q} = \sum_{\alpha} q_{\alpha} \hat{e}_{\alpha}$ is small) 
satisfy the general relation \cite{KupcicPRB2,Wooten}
\begin{eqnarray}
&&eq^{L'L} ({\bf k}+{\bf q},{\bf k})  
\approx  e \delta_{L,L'} 
\label{eq12} \\ \nonumber  
&& \hspace{20mm} 
+ \big(1-\delta_{L,L'} \big) \sum_{\alpha} 
\frac{\hbar q_{\alpha} J_{\alpha}^{L'L} ({\bf k})}{ 
E_{L'}({\bf k}+{\bf q}) - E_{L}({\bf k})},
\end{eqnarray}
%
%
with the longitudinal current vertices $J_{\alpha}^{L'L} ({\bf k})$ 
identical to the transverse  current vertices given by Eqs.~(\ref{eq10}).

\section{Raman correlation functions in pure systems}
%
%

In the mean-field slave-boson theory \cite{Kotliar} used here,
the physical Raman correlation functions
are proportional to the corresponding correlation functions of the auxiliary
fermions described by the band structure associated with
Eqs.~(\ref{eq4}) and (\ref{eq5}). 
 	It goes without saying that the same conclusions hold for the physical 
fermions with the negligible local interactions $U_{\rm d}$.
	The simplest operative way to determine the Raman correlation functions 
of this three-band auxiliary fermion  model
is to consider the Goldstone theorem for the
thermodynamic potential
in the Matsubara representation with $H' = H^{\rm ext} +H_2'+ H_1'$ 
representing the perturbation, and collect all fourth-order contributions
in the vector fields $A_{\alpha} ({\bf q}'')$ and $A_{\beta} ({\bf q}')$.
	It is convenient to divide this procedure 
into four steps.
	First, the $H' = H^{\rm ext}$ case provides the definition of the Raman
vertex functions in the multiband model under consideration,
with  particular care devoted to the resonant enhancement 
of the Raman scattering processes. 
	Second, for $H' = H^{\rm ext} +H_{\rm c}$, we shall define the direct 
contributions to the Raman correlation functions and reconsider the role 
of the long-range screening in the pure multiband models.
	Third, by considering the perturbation 
$H' = H^{\rm ext} +H_{\rm c}+ H_1'$, we shall introduce the 
distinction between the direct and indirect (disorder-assisted)
electron-hole excitations and discuss  which of these 
processes dominate the Raman spectra measured in the high-$T_c$ cuprates.
	Finally, by including $H_{\rm AF}$, we shall study the influence
of the low-frequency excitations across the AF (pseudo)gap on both
the Drude part and the related 
low-lying interband part of the Raman spectrum.

\subsection{Raman vertex functions in pure systems}
%
%

In the absence of the disorder and AF scattering processes, 
the direct summation of the fourth-order diagrams
in the vector fields $A_{\alpha} ({\bf q}'')$ and $A_{\beta} ({\bf q}')$
leads to  Fig.~1(a), 
representing the Raman  correlation function in the ideal lattice,
approximately given by  its intraband contribution.
	Namely, in the high-$T_c$ cuprates, the interband excitation energies 
are of the order of typical optical energies, \hbox{1.75--2.75} eV,
which is far above the largest Raman shift (defined below) 
measured in experiments ($\hbar \omega < 1$ eV).
	Consequently, the interband contributions to 
the Raman  correlation functions  can safely be neglected in the 
ideal lattice.
	As will be seen below, the AF correlations
introduce the possibility of the low-lying ``interband'' 
excitations requiring the generalization (Sec. III B) of Fig.~1(a).

Thus, in a pure system (denoted by  $p$) we have 
\begin{eqnarray} 
&& \chi_{\alpha \beta, \beta \alpha}^{p}
({\bf q}, \omega, \omega_{i}) 
\approx \frac{1}{ v}  \sum_{{\bf k} {\bf k}' \sigma} 
\gamma^{cc} _{\alpha \beta} ({\bf k}, \omega_{i}, \omega_{s})
\label{eq13} \\ \nonumber  
&&  \hspace{20mm}
\times \frac{1}{\hbar } 
{\cal D}^{cc}_{p} ({\bf k}, {\bf k}_+,{\bf k}'_+, {\bf k}',  \omega) 
\gamma^{cc} _{\beta \alpha} ({\bf k}', \omega_{s}, \omega_{i}), 
\end{eqnarray}
%
%
where
${\cal D}^{cc}_{p} ({\bf k}, {\bf k}_+ ,{\bf k}'_+, {\bf k}',  \omega)$ 
is the  intraband electron-hole propagator in the ideal lattice,  
defined by 
\begin{eqnarray}
 && \frac{1}{\hbar} 
{\cal D}^{LL'}_{p}  ({\bf k}, {\bf k}_+,{\bf k}'_+, {\bf k}', \omega) 
\\  \nonumber 
 && 
 \hspace{15mm} =
 \delta_{{\bf k}, {\bf k}'}
\frac{f_L({\bf k}) - f_{L'}({\bf k}+ {\bf q})}{\hbar \omega 
+ E_L({\bf k}) - E_{L'}({\bf k}+ {\bf q}) + {\rm i} \eta},
\label{eq14} 
\end{eqnarray}
%
%
for the band indices $L = L' =c$.
	$f_L({\bf k}) \equiv f(E_L({\bf k}))$ is the Fermi--Dirac 
distribution function.
	Furthermore,  the 
$\gamma^{cc} _{\alpha \beta} ({\bf k}, \omega_{i}, \omega_{s})$
are the related intraband Raman vertices 
\begin{eqnarray}
&&\gamma^{cc} _{\alpha \beta} ({\bf k}, \omega_{i}, \omega_{s} ) =
- \frac{m}{e^2}  
\sum_{L \neq c} \bigg[
\frac{J_{\alpha}^{Lc} ({\bf k}) J_{\beta}^{cL} ({\bf k})}{
\hbar \omega_{i}   - E_{Lc} ({\bf k})  + i \eta  } 
\nonumber \\
&& \hspace{10mm}
- \frac{J_{\alpha}^{cL} ({\bf k}) J_{\beta}^{Lc} ({\bf k})}{
\hbar \omega_{s}   + E_{Lc} ({\bf k}) + i \eta   } \bigg] 
+ \gamma^{cc} _{\alpha \beta} ({\bf k} ;2),
\label{eq15} 
\end{eqnarray}
%
%
and ${\bf k}_+ = {\bf k}+ {\bf q}$.
	Here $\omega_{i}, {\bf q}'', \alpha$ and  
$\omega_{s}, {\bf q}', \beta$ are the frequencies, wave vectors
and polarization indices of the incoming and scattered photons,
respectively.
	$\omega = \omega_{i} - \omega_{s}$ is the Raman shift, 
${\bf q} = {\bf q}'' - {\bf q}'$, and  
$ E_{LL'} ({\bf k}) = E_{L} ({\bf k}) -E_{L'} ({\bf k})$. 
	Eq.~(\ref{eq15}) is gauge invariant in the limit 
$\eta \rightarrow 0$.
	As mentioned at the beginning of this section,
both the scattering processes on the disorder and the AF correlations 
are absent in $\gamma^{cc}_{\alpha \beta} ({\bf k}, \omega_{i}, \omega_{s})$.

   \begin{figure}[tb]
     \includegraphics[height=10pc,width=20pc]{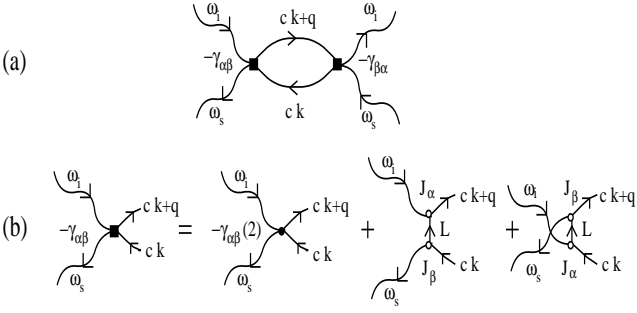}
    \caption{
    (a) The  purely electronic intraband Raman correlation functions 
    in a pure system.
    (b) The  Raman vertex  (full rectangle)
    shown in terms of the bare Raman vertex (full circle) and the
    interband current vertices (open circles).
    }
    \end{figure}

The diagrammatic representation of the Raman vertices is shown
in Fig.~1(b).
	The first term on the right-hand side is the quadratic coupling 
term, while the latter two represent the bi-linear contributions.
	The resonant nature of the Raman scattering processes refers to
the bi-linear terms. 
	The resonant effects are  large in 
the high-$T_c$ cuprates because, as mentioned above,  
the interband excitation energies $E_{Lc} ({\bf k})$ in Eq.~(\ref{eq15})
are of the order of typical optical energies.
	In addition to the resonant condition, 
$E_{Lc} ({\bf k}) \approx \hbar \omega_{i}$
and/or $E_{Lc} ({\bf k}) \approx  \hbar \omega_{s}$, 
the efficiency of the resonant enhancement of the Raman scattering
processes depends also on the relaxation processes in the intermediate 
interband photon absorptions/emissions that are omitted here.
	Although, in principle, these relaxation processes have 
to be treated on an equal footing with the relaxation processes 
in the electron-hole propagators
${\cal D}^{LL'} ({\bf k}, {\bf k}_+, {\bf k}'_+,{\bf k}', \omega)$, 
we shall use below an approximate treatment,
by including the former  phenomenologically (see Sec.~III A.2)
and the latter by using the direct summation method (Sec.~IV).

\subsubsection{Effective mass theorem}

Let us consider  the
$\omega_{i} = \omega_{s} = 0$, $\eta \rightarrow 0$ limit of 
Eq.~(\ref{eq15}).  
	The result is  the static Raman vertex of the form
\begin{equation}
\gamma^{cc} _{\alpha \beta} ({\bf k}) =
\gamma^{cc} _{\alpha \beta} ({\bf k} ;2) +\frac{m}{e^2}  
\sum_{L \neq c} 
\frac{2 J_{\alpha}^{Lc} ({\bf k}) J_{\beta}^{cL} ({\bf k})
}{E_{Lc} ({\bf k})  }. 
\label{eq16}
\end{equation}
%
%
	Here  the symmetry relation
$J_{\alpha}^{Lc} ({\bf k}) = J_{\alpha}^{cL} ({\bf k})$
has been used.
	This expression can be combined with the relation
\begin{eqnarray}
\gamma^{cc} _{\alpha \beta} ({\bf k}) &=& 
\mp \frac{m}{\hbar^2}
 \frac{\partial^2 E_c ({\bf k}) }{ \partial k_{\alpha}\partial k_{\beta}} 
\nonumber
\end{eqnarray}
to obtain the ``effective mass'' theorem 
\begin{eqnarray}
\mp  \frac{m}{\hbar^2}
 \frac{\partial^2 E_c ({\bf k}) }{ \partial k_{\alpha}\partial k_{\beta}}&=&
\gamma^{cc} _{\alpha \beta} ({\bf k} ;2) 
\nonumber \\  
 &&
+\frac{m}{e^2}  \sum_{L \neq c} 
\frac{2 J_{\alpha}^{Lc} ({\bf k}) J_{\beta}^{cL} ({\bf k})
}{E_{Lc} ({\bf k})  }.
 \label{eq17} 
\end{eqnarray}
%
%
	Eq.~(\ref{eq17}) [and Eq.~(\ref{eq19})] holds even when its 
left-hand side is dependent on {\bf k}, i.e. beyond the effective mass
approximation in the vicinity of the Fermi level.
	The result  is appropriate for any multiband model 
with the hole-like ($-$ sign, the case 
considered here) or electron-like ($+$ sign)  dispersion of the 
conduction electrons.

Eq.~(\ref{eq17})  turns out to be  important  
for both the conductivity-sum-rule 
analyses and  the transport-coefficient studies, 
in particular when  the AF  term (\ref{eq5}) is included.
	Actually, Eq.~(\ref{eq17}) represents a partial conductivity sum rule 
for three bands \cite{KupcicPC}, which holds when the photon frequencies 
are small with  
respect to the transition frequencies  into all other bands.
	When the high-frequency transitions  are included 
in the present approach
the ``effective mass'' is replaced by the free carrier mass, i.e.
the present tight-binding (Wannier) approach \cite{ZBB}
satisfies the general sum rule established by Abrikosov 
and Genkin \cite{Abrikosov,Kosztin}.

The theorem states that the zero-frequency  electron-hole pairs 
(corresponding to the formal limit  $\omega_{i}, \omega_{s} \rightarrow 0$) 
can be excited by the electromagnetic fields through the bare quadratic
electron-photon coupling and/or through  the bi-linear term  in which 
the first-order (high-frequency) interband excitations appear as virtual
intermediate states.

\subsubsection{Elastic-Raman-vertex approximation}

Since the  Raman shift $\omega = \omega_{i} - \omega_{s}$ is small in
comparison with the typical values of $\omega_{i}$ or $\omega_{s}$,
it is reasonable, in the numerical calculation in Sec.~V,
to use the elastic-Raman-vertex approximation 
\begin{eqnarray}
\gamma^{cc} _{\alpha \beta} ({\bf k}, \omega_{i}, \omega_{s} ) 
&\approx&
\gamma^{cc} _{\alpha \beta} ({\bf k},\omega_{i}, \omega_{i} )
\equiv 
\gamma^{cc} _{\alpha \beta} ({\bf k},\omega_{i}),
\label{eq18}
\end{eqnarray}
%
%
in which the zero-frequency processes 
($\omega_{i}, \omega_{s} \approx 0$)
are approximately  separated from the  higher-frequency absorption/emission 
processes.
	The phenomenological treatment of the interband relaxation processes 
in the resonant channel then gives rise to the general gauge-invariant  
expression which reduces to Eq.~(\ref{eq15}) in the limit
$\Gamma^{\rm inter}/\omega_{i} \rightarrow 0$ 
\begin{eqnarray}
\gamma^{cc} _{\alpha \beta} ({\bf k}, \omega_{i} ) &=& 
\gamma^{cc} _{\alpha \beta} ({\bf k} ) 
- \frac{m}{e^2}  \sum_{L \neq c} 
\frac{(\hbar \omega_{i})^2
J_{\alpha}^{Lc} ({\bf k}) J_{\beta}^{cL} ({\bf k})
}{E^2_{Lc} ({\bf k}) } 
\nonumber \\ 
&& \times \frac{2E_{Lc} ({\bf k})}{
(\hbar \omega_{i}  + i \hbar \Gamma^{\rm inter} )^2 - E^2_{Lc} ({\bf k})  } 
\label{eq19}
\end{eqnarray}
%
%
[again $J_{\alpha}^{Lc} ({\bf k}) = J_{\alpha}^{cL} ({\bf k})$ is used].

It is useful now to incorporate the symmetry properties
of the Emery three-band model into Eqs.~(\ref{eq15}) and (\ref{eq19}).
	First, we remember that the analysis of the
electronic Raman spectra of the high-$T_c$ cuprates is usually focussed on 
the in-plane polarization of the electromagnetic fields
($\alpha, \beta = x, y$).
	It is thus convenient to arrange the  Raman 
vertices according to the irreducible representations of the $D_{4h}$ point 
group \cite{BarisicIJMP,Barisic,Devereaux1}.
	The resulting Raman vertices are of the form 
$\gamma_{\nu  }^{cc} ({\bf k}, \omega_{i} )$, with the label  
$\nu = A_{1g},$ $B_{1g}$, and $B_{2g}$
representing the $A_{1g}$, $B_{1g}$, and $B_{2g}$ Raman channels, respectively.
	The symmetrized vertices are
\begin{eqnarray}
\gamma^{cc} _{A_{1g}} ({\bf k}, \omega_{i} ) &=&
\gamma^{cc} _{xx} ({\bf k}, \omega_{i} ) + 
\gamma^{cc} _{yy} ({\bf k}, \omega_{i} ), \nonumber \\
\gamma^{cc} _{B_{1g}} ({\bf k}, \omega_{i} ) &=&
\gamma^{cc} _{xx} ({\bf k}, \omega_{i} ) - 
\gamma^{cc} _{yy} ({\bf k}, \omega_{i} ), \nonumber \\
\gamma^{cc} _{B_{2g}} ({\bf k}, \omega_{i} ) &=&
\gamma^{cc} _{xy} ({\bf k}, \omega_{i} ). 
\label{eq20}
\end{eqnarray}
%
%
	It should be noticed here that the Raman correlation functions 
of the tetragonal high-$T_c$ cuprates
are  diagonal in this  representation.
	The orthorhombic distortion of the CuO$_2$ plane, 
which occurs in some  compounds  (YBa$_2$Cu$_3$O$_{7-x}$, for example),
mixes these three  channels.
	However, as previously estimated \cite{KupcicPRB2}, the mixing 
is typically of the order of 1/10 and is  neglected 
in the present analysis.

\subsection{Long-range screening in pure systems}

\begin{figure}[tb]
\includegraphics[height=8pc,width=18pc]{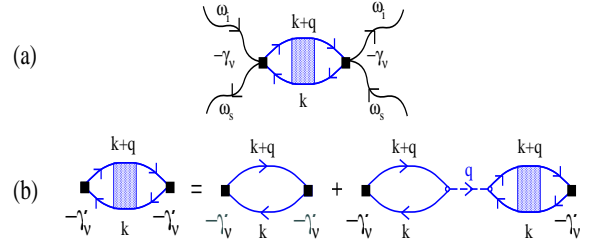}
\caption{
    (a) The  Raman correlation functions in a general case
    with the long-range forces and the quasi-elastic scattering processes
    taken into account. 
    The full rectangle is the Raman vertex of Fig.~1(b). 
     The shaded box includes the  electron-hole self-energy contributions
    associated with both the long-range forces 
    and the scattering processes on the disorder.             
    (b)  The long-range screening of the  Raman correlation functions
    in the case where the scattering processes on the disorder are absent.  
    The open circles represent the charge vertices and the 
    dashed line is the long-range force $4\pi e^2/q^2$.
    }
\end{figure}

	The effects of the long-range Coulomb forces on 
the Raman correlation functions are given in the usual way 
\cite{Klein,Monien,Ruvalds,Devereaux2,BarisicSSC,KupcicPRB2,Sherman2}. 
	In absence of the incoherent scattering processes those
functions are described by the diagrams in  Fig.~2(b).
	The screened correlation function 
$ \widetilde{\chi}_{\nu,\nu} ({\bf q}, \omega, \omega_i)$ is given by
\begin{eqnarray}
&& \widetilde{\chi}_{\nu,\nu}({\bf q}, \omega, \omega_i) =
\chi_{\nu,\nu} ({\bf q}, \omega, \omega_i) 
\label{eq21}  \\  \nonumber 
&& \hspace{20mm} 
+ \chi_{\nu, 1} ({\bf q}, \omega, \omega_i)
\frac{4 \pi e^2}{ q^2  \varepsilon({\bf q}, \omega)}
\chi_{1, \nu} ({\bf q}, \omega, \omega_i).
\end{eqnarray}	
%
%
	The coupling function 
$\chi_{\nu, 1} ({\bf q}, \omega, \omega_i)$ 
is defined by Eq.~(\ref{eq13}), with 
$\gamma^{cc} _{\alpha \beta} ({\bf k}, \omega_{i})
\gamma^{cc} _{\beta \alpha } ({\bf k}', \omega_{i})$
replaced by 
$\gamma^{cc} _{\nu} ({\bf k}, \omega_{i})  
q^{cc}({\bf k'}+{\bf q},{\bf k'})$.
	The dielectric function in Eq.~(\ref{eq21}) 
has the form
\begin{eqnarray} 
\varepsilon ({\bf q}, \omega) &=& \varepsilon_{\infty} ({\bf q}, \omega) 
- \frac{4 \pi e^2}{q^2 } \chi_{1, 1} ({\bf q}, \omega),
\label{eq22} 
\end{eqnarray}
%
%
with $e^2 \chi_{1, 1} ({\bf q}, \omega)$ representing 
the  charge-charge correlation function given by
\begin{eqnarray}
&& e^2 \chi_{1, 1} ({\bf q}, \omega) =  
\frac{1}{v}  \sum_{LL'} \sum_{{\bf k} {\bf k}' \sigma} 
e^2 q^{LL'}  ({\bf k}, {\bf k}+ {\bf q}) 
\label{eq23}   \\ \nonumber  
&& 
\hspace{15mm} \times
q^{L'L}  ({\bf k}'+ {\bf q},{\bf k}')
\frac{1}{\hbar } 
{\cal D}^{LL'} ({\bf k}, {\bf k}_+, {\bf k}'_+, {\bf k}', \omega).
\end{eqnarray}
%
%
	Here 
${\cal D}^{LL'} ({\bf k}, {\bf k}_+, {\bf k}'_+, {\bf k}', \omega)$
is the electron-hole propagator defined in Appendix C.

For  the $B_{1g}$ and  $B_{2g}$ Raman channels, the coupling functions
$\chi_{\nu, 1} ({\bf q},\omega, \omega_i)$ 
vanish for  symmetry reasons, and the long-range forces do not affect
the  Raman spectra in the $B_{1g}$ and  $B_{2g}$ channels.
	Furthermore, it is useful to separate 
the constant term in the $A_{1g}$ Raman vertex from the dispersive term
\cite{Abrikosov,Kosztin}, 
$\gamma^{cc} _{A_{1g}} ({\bf k}, \omega_{i}) =
\overline{\gamma}^{cc} _{A_{1g}} (\omega_{i})+
\hat{\gamma}^{cc} _{A_{1g}} ({\bf k},\omega_{i})$,
in the way that 
$\hat{\chi}_{A_{1g}, 1}({\bf q},\omega, \omega_i)=0$
[notice that $\gamma^{cc} _{\nu} ({\bf k},\omega_i)= 
\hat{\gamma}^{cc} _{\nu}({\bf k},\omega_i)$ 
for $\nu = B_{1g},B_{2g}$].
	In this way 
$\hat{\chi}_{\nu, 1} ({\bf q},\omega, \omega_i)=0$
for all three Raman channels.
	[The hat in $\hat{\chi}_{\nu, 1}({\bf q},\omega, \omega_i)$
indicates that only the dispersive part of the vertex 
$\gamma^{cc} _{\nu}({\bf k},\omega_i)$,
$\hat{\gamma}^{cc} _{\nu}({\bf k},\omega_i)$, 
is included in 
$\chi_{\nu, 1}({\bf q},\omega, \omega_i)$.]
 	Consequently, the dispersive terms 
$\hat{\gamma}^{cc} _{\nu} ({\bf k},\omega_{i})$ 
are unaffected by the long-range screening,
at least in pure systems, while the constant term 
$\overline{\gamma}^{cc} _{A_{1g}} (\omega_{i})$
is screened in the same way as the monopole charge
$q^{cc} ({\bf k},{\bf k}+ {\bf q}) \approx 1$
\cite{Pines,Mahan,Ziman2}.

The Raman spectra, associated with imaginary part of Eq.~(\ref{eq21}),
comprise the incoherent electron-hole contributions characterized by 
the cut-off frequency of the order of $qv_{\rm F}$
and, for the $A_{1g}$ channel, by the plasmon contribution
related to the screening of 
$\overline{\gamma}^{cc} _{A_{1g}} (\omega_{i})$.
	These spectra are directly related to the dynamical 
structure factor 
$S({\bf q}, \omega) = 
- {\rm Im}\{ \widetilde{\chi}_{1,1}({\bf q}, \omega) \}$.	
	The intensity of both the collective and incoherent electron-hole 
contributions to 
$-{\rm Im }\{\widetilde{\chi}_{\nu,\nu}({\bf q}, \omega, \omega_i) \}$
is proportional to small $q^2$. 
	These types of signals have never been
detected in the high-$T_c$ cuprates \cite{Uchida,Ruvalds},
in contrast to  the semiconducting systems, such as GaAs
($qv_{\rm F} \approx 50$ cm$^{-1}$)
\cite{Platzman}.
	In the high-$T_c$ cuprates, the measured Raman
spectra are roughly proportional to the optical conductivity,
with the intensity proportional to the channel-dependent relaxation rates. 	
	This leads us to study the scattering of the quasi-particles on 
the disorder.

\section{Raman correlation functions in systems with disorder}
%
%
\subsection{Incoherent scattering }
%
%

This section deals with the contributions of the incoherent 
quasi-elastic scattering  to the Raman correlation functions
$\widetilde{\chi}_{\nu,\nu} ({\bf q}, \omega, \omega_i)$,
including  the Coulomb screening effects.
The discussion starts from the low order scattering on the disorder,
continues by the summations to high orders and adds the Coulomb 
screening at the end.
	In this discussion it is convenient to distinguish between 
the direct and indirect processes, as further explained below.

\subsubsection{Direct processes}

   \begin{figure}[tb]
    \includegraphics[height=6pc,width=16pc]{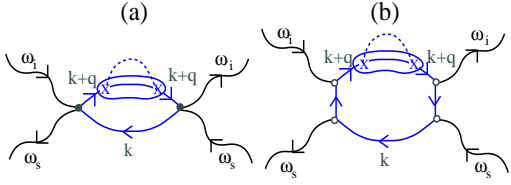}    
    \caption{
     Two typical quadratic (a) and bi-linear (b) direct Raman scattering processes  
    in the conduction band proportional to $(H_1')^2$.
    The self-energy parts on the diagrams treated  as constant
    are encircled \cite{KupcicFA}.    
    The crosses represent the quasi-elastic scattering $H_1'$.
    }
    \end{figure}

As illustrated in Fig.~3,
for all correlation functions considered in this article
(charge-charge, current-current and Raman correlation functions),
the probability for the direct 
electron-hole pair creation is proportional to 
$f_c({\bf k}) - f_c({\bf k}+{\bf q})$ and associated with the resonance
condition $\hbar \omega \approx E_c({\bf k}) - E_c({\bf k}+{\bf q})$.
	The corresponding scattering paths $1 \rightarrow 3$
and $1 \rightarrow 2 \rightarrow  3$ are  shown in Fig.~4. 
	The direct  scattering on the disorder can be roughly incorporated
in the correlation functions in the standard phenomenological way \cite{Ziman2}.
	Alternatively, one can apply the gauge-invariant
treatment  to sum  the direct processes shown in  Fig.~5  
in powers of  $(H_1')^2$. 
	The gauge invariance conserves the number of charge carriers
in the scattering processes \cite{Pines}.
	As shown in Appendix C, for $\omega > qv_{\rm F}$, 
the latter approach gives
the unscreened, direct  charge-charge correlation function 
(intra- and interband contributions) of the form
\begin{eqnarray}
&& \! \! \! \! \! \! \! \! 
e^2\chi_{1,1}^{\rm d} ({\bf q}, \omega) = 
\frac{1}{v} \sum_{\alpha' LL' {\bf k} \sigma}
\frac{q_{\alpha'}^2 }{\omega^2} 
\bigg( \frac{ \hbar \omega}{E_{L'L}({\bf k}_+,{\bf k})} \bigg)^{n_{LL'}} 
\! \! \! \!
\big| J_{\alpha'}^{LL'} ({\bf k}) \big|^2
 \nonumber \\ 
&&   \times
\frac{f_L({\bf k})-f_{L'}({\bf k}_+) }{
\hbar\omega + {\rm i}\hbar \Gamma_{\alpha'}^{LL'}({\bf k},\omega)  
+ E_{LL'}({\bf k},{\bf k})
-\displaystyle \frac{E_{L'L'}^2({\bf k},{\bf k}_+)}{\hbar \omega}},
 \nonumber \\ 
\label{eq24} 
\end{eqnarray}
%
%
where ${\bf q} = \sum_{\alpha'} q_{\alpha'} \hat{e}_{\alpha'}$,
$n_{LL} =1$, $n_{L\underline{L}} =2$,
$\Gamma_{\alpha'}^{LL'}({\bf k},\omega) =
{\rm Im} \{\Sigma_{\alpha'}^{LL'}({\bf k},\omega)\}$
and
$E_{LL'}({\bf k},{\bf k}_+) = E_{L}({\bf k})-E_{L'}({\bf k}_+)$.

   \begin{figure}[tb]
    \includegraphics[height=12pc,width=17pc]{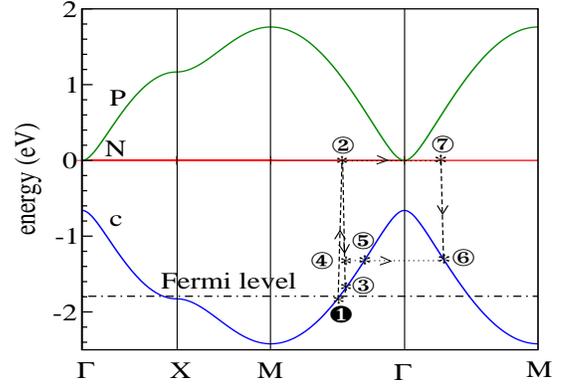}    
    \caption{The direct ($1 \rightarrow 2 \rightarrow  3$) and
    indirect (forward, $1 \rightarrow 2 \rightarrow  4 \rightarrow 5$,
    or backward, $1 \rightarrow 2 \rightarrow  4 \rightarrow 6$)
    bi-linear  Raman scattering processes in the conduction band.   
    The solid lines represent the three effective fermionic bands 
    (the indices $c$, $P$ and $N$) 
    for the typical values of the model parameters
    $\Delta^{\rm eff}_{\rm pd} = 0.66$ eV and 
    $t^{\rm eff}_{\rm pd} = 0.73$ eV
    \cite{KupcicPC,KupcicPRB1}.
    	The energies are measured with respect to the energy of the 
   $2p_{\sigma}$ oxygen orbitals, so that the dispersionless nonbonding 
    band is placed at $E_{\rm p} = 0$.
    The dashed lines are the photon dispersions, and
    the dot-dashed line is the Fermi energy $\mu = -1.793$ eV 
    corresponding to the hole doping $\delta = 0.1$.
    }
    \end{figure}
\begin{figure}[tb]
\includegraphics[height=3.pc,width=18pc]{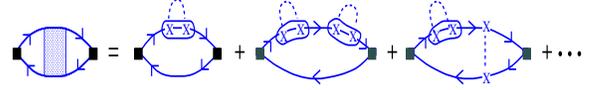}
\caption{
 	A few direct contributions to the Raman correlation functions
	in powers of $(H_1')^2$, according to Eqs.~(\ref{eqC9}), (\ref{eqC8})
	and (\ref{eqC4}).  
	The full rectangle is the effective Raman vertex of Fig.~1(b).
}
\end{figure}

Eq.~(\ref{eq24}) can be easily generalized to other correlation functions.
	For the quasi-elastic scattering 
${\rm Im} \{\Sigma^{cc}({\bf k},\omega)\} \approx \Gamma^{c,{\rm d}}_i$ 
(here, the index $i =1$, $\alpha$, and $\nu$ for the charge, current, 
and Raman vertices, respectively).
 	In the dynamical limit, we thus obtain   
the universal expression for the unscreened, direct intraband 
correlation functions
\begin{equation}
\chi_{i,j}^{\rm d} ({\bf q}, \omega) = 
\sum_{\alpha'}
\frac{q_{\alpha'}^2}{ \omega} \frac{1}{\omega + {\rm i} \Gamma^{c,{\rm d}}_i}
\frac{(at_{\rm pd}^{\rm eff})^2}{v_0\hbar^2} 
n_{i,j}^{\rm d} (\mu).
\label{eq25}
\end{equation}
%
%
	Here $n_{1,1}^{\rm d} (\mu)$ is the effective density of states
at the Fermi energy given by
\begin{equation}
n_{1,1}^{\rm d}  (\mu)  = -\frac{1}{N}  
\sum_{{\bf k} \sigma}  
\big| q^{cc} ({\bf k},{\bf k}+{\bf q})  j_{\alpha'}^{cc} ({\bf k})\big|^2
\frac{\partial f_c ({\bf k})}{\partial E_c ({\bf k})},
\label{eq26}
\end{equation}
%
%
while  $n_{\alpha, \alpha}^{\rm d} (\mu)$,  $n_{\nu,\nu}^{\rm d} (\mu)$
and $n_{\nu,1}^{\rm d} (\mu)$
are obtained by replacing 
$\big| q^{cc} ({\bf k},{\bf k}+{\bf q}) \big|^2 = 1$ in Eq.~(\ref{eq26}) 
with $\big( j_{\alpha  }^{cc} ({\bf k}) \big)^2$,
$\big| \gamma_{\nu  }^{cc} ({\bf k}, \omega_{\rm i}) \big|^2$,
and $\gamma_{\nu  }^{cc} ({\bf k}, \omega_{\rm i}) 
q^{cc}({\bf k}+{\bf q}, {\bf k})$,
respectively.
	Finally, $j_{\alpha  }^{cc} ({\bf k}) = 
\hbar J_{\alpha}^{cc} ({\bf k})/(eat_{\rm pd}^{\rm eff})$ 
is the dimensionless current vertex,  Eq.~(\ref{eqA5}), and 
$v_0$ is the unit cell volume.
	For the electromagnetic fields ($i = \alpha, \nu$)
the wave vector ${\bf q} = \sum_{\alpha'} q_{\alpha'} \hat{e}_{\alpha'}$ 
is perpendicular to the polarization of the fields;
i.e. $q_{\alpha'} = q_z$ for the symmetrized Raman vertices in
Eq.~(\ref{eq20}).

The  RPA series for the screened direct contribution to the Raman
correlation functions is illustrated in Fig.~6(a), and is given by inserting 
the expression (\ref{eq25}) into Eq.~(\ref{eq21}).
	As can be easily seen, the intensity of both the plasmon and 
electron-hole incoherent  contributions to 
$-{\rm Im} \{ \widetilde{\chi}_{\nu,\nu}^{\rm d} 
({\bf q}, \omega, \omega_i) \}$ remains proportional to small $q^2$.
	Fig.~6(b) represents the quadrupolar analog 
of the well-known Hopfield series \cite{Mahan}.
 	It will be argued below that the latter is not important 
for the Raman scattering on the high-$T_c$ superconductors.

\begin{figure}[tb]
\includegraphics[height=7pc,width=18pc]{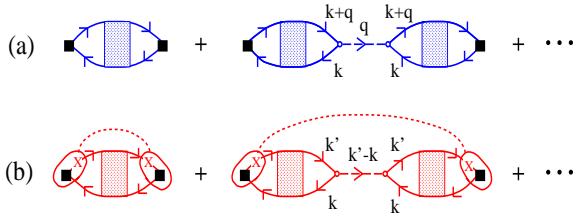}
\caption{
    The Coulomb screening of the direct (a) and indirect (b) 
    processes in the Raman response functions 
    in presence of the quasi-elastic scattering.   
    The dotted box includes the electron-hole 
    self-energy contributions associated  with the quasi-elastic scattering 
    processes.
}
\end{figure}

   \begin{figure}[tb]
    \includegraphics[height=6pc,width=20pc]{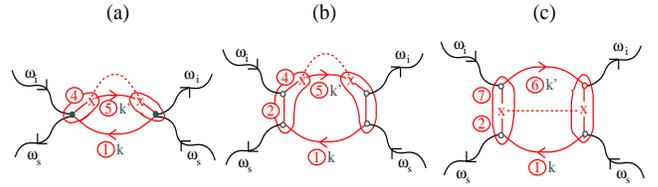}
        
    \caption{
    Three typical indirect Raman scattering processes
    proportional to $(H_1')^2$.
    The first two include the incoherent scattering of conduction electrons   
    while the third shows the incoherent scattering in the empty band(s)
    (the notation is the same as in Fig.~4).
    The effective vertices  are encircled \cite{KupcicFA}. 
       }
    \end{figure}

\subsubsection{Indirect processes }

Omitting again the Coulomb screening to begin with,
the disorder-assisted,  indirect electron-hole contribution 
is associated to $f_c({\bf k}) - f_c({\bf k}')$,
with uncorrelated ${\bf k}$ and ${\bf k}'$ 
[see the $1 \rightarrow 4 \rightarrow 5$ processes  shown in Fig.~4
and the related diagrams in Fig.~7(a), as well as 
the $1  \rightarrow 2 \rightarrow 4 \rightarrow 5$ processes 
represented by the diagram in Fig.~7(b)].
	These types of processes become important when
$\hbar \omega \gg |E_c({\bf k}) - E_c({\bf k}+{\bf q})|$,
with the resonance at $\hbar \omega \approx E_c({\bf k}) - E_c({\bf k}')$.
	This is a typical situation encountered in the absorption
and/or emission of photons by conduction electrons,
i.e. in the intraband optical-conductivity and Raman experiments on metals.
	On the other hand, the indirect Raman scattering processes 
$1  \rightarrow 2 \rightarrow 7 \rightarrow 6$,
shown in Fig.~7(c), are directly related to the indirect interband 
optical conductivity \cite{Ziman2}.
 	For the time-dependent $H_1'$ they are 
essential for the Raman analysis of the
insulating and semiconducting systems \cite{Cardona}.
	In the present case, $H_1'$ includes only the quasi-elastic scattering 
and therefore the diagram in Fig.~7(c) has the resonant behavior 
similar to the diagram in  Fig.~7(b).
	Thus the processes in Fig.~7(c) can be included in the effective 
Raman vertex (\ref{eq19}) and will not be discussed hereafter.

   \begin{figure}[tb]
    \includegraphics[height=8pc,width=20pc]{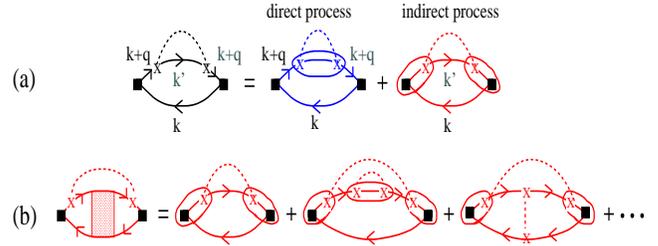}    
    \caption{
    (a) The direct and indirect high-frequency contributions 
    [proportional to $(H_1')^2$] to the Raman correlation functions
    [Fig.~3 and Fig.~7(a,b)].
    (b) A few indirect  leading terms in powers 
    of $(H_1')^2$. 
    }
    \end{figure}

The direct and indirect scattering processes, shown in Fig.~8(a),  
are large  in the high-frequency limit [$\propto (H_1')^2$].
	The first qualitatively important corrections 
to the indirect high-frequency term
come from the second and third term 
in Fig.~8(b) which are proportional to $(H'_1)^4/\omega$,
i.e. they are singular in the zero-frequency limit.
	The consistent treatment of the indirect Raman scattering processes
requires thus  the summation to infinity
of the most singular terms in powers of $(H_1')^2/\omega$.
	This requires summing the  singular
contributions to all orders in $(H_1')^2/\omega$
in order to obtain the description 
which is correct in both the high- and low-frequency limits.

As explained  in Ref.~\cite{KupcicFA} in the example 
of optical conductivity,
the gauge-invariant treatment of the single-particle self-energy 
and vertex corrections in the  indirect processes gives rise to
effective vertices in which there is a complete cancellation
of the scattering processes associated with the constant 
terms in the bare vertices.
	In the case of optical conductivity, this means 
that the indirect processes in the charge-charge
correlation functions  are absent altogether because the effective vertex
$[q^{cc}({\bf k},{\bf k})- q^{cc}({\bf k}',{\bf k}')]
V^{cc}_1({\bf k}- {\bf k}')/(\hbar \omega)$
vanishes due to the fact that $q^{cc}({\bf k},{\bf k}) \approx 1$.
	In the Raman case, the effective vertices 
$[\gamma^{cc} _{\nu} ({\bf k}, \omega_{i})
- \gamma^{cc} _{\nu} ({\bf k}', \omega_{i})]
V^{cc}_1({\bf k}- {\bf k}')/(\hbar \omega)$
in $\chi^{\rm id}_{\nu,\nu} (\omega, \omega_i)$	
are given by the sum of two terms shown in Fig.~9(a), 
setting 
$\gamma^{cc} _{\nu} ({\bf k}, \omega_{i}) = 
\overline{\gamma}^{cc} _{\nu} (\omega_{i}) 
+ \hat{\gamma}^{cc} _{\nu} ({\bf k}, \omega_{i})$,
and  reduce to $[\hat{\gamma}^{cc} _{\nu} ({\bf k}, \omega_{i})
- \hat{\gamma}^{cc} _{\nu} ({\bf k}', \omega_{i})]
V^{cc}_1({\bf k}- {\bf k}')/(\hbar \omega)$.
	The contribution to 
$\chi^{\rm id}_{\nu,\nu} (\omega, \omega_i)$
of the constant terms 
$\overline{\gamma}^{cc} _{\nu} (\omega_{i})$,
present only in the $\nu = A_{1g}$ channel, thus vanishes, 
in analogy with the case of optical conductivity.
	In this way,
$\chi^{\rm id}_{\nu,\nu} (\omega, \omega_i) =
\hat{\chi}^{\rm id}_{\nu,\nu} (\omega, \omega_i)$ with the hat 
again indicating that only  the dispersive terms 
$\hat{\gamma}^{cc} _{\nu} ({\bf k}, \omega_{i})$ in the Raman vertices
contribute to $\chi^{\rm id}_{\nu,\nu} (\omega, \omega_i)$.

   \begin{figure}[tb]
    \includegraphics[height=14pc,width=20pc]{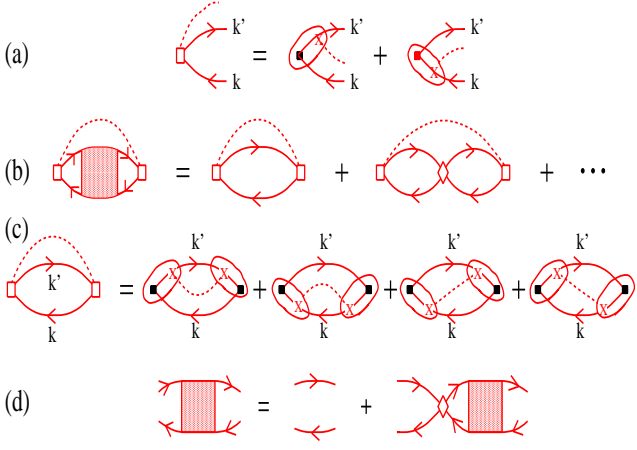}    
    \caption{
    (a) The effective Raman vertex (open rectangle) in the indirect processes,
    $[\hat{\gamma}^{cc} _{\nu} ({\bf k}, \omega_{i})
    - \hat{\gamma}^{cc} _{\nu} ({\bf k}', \omega_{i})]
     V^{cc}_1({\bf k}- {\bf k}')/(\hbar \omega)$.
    (b) The expansion of the indirect contribution to the Raman correlation
    functions in powers 
    of $(H_1')^2/\omega$, with the leading term explicitly shown in (c).
    The shaded box is the electron-hole propagator which is obtained by 
    the self-consistent solution of the equation shown in (d) \cite{KupcicFA}.
    The diamond is the electron-hole self-energy containing both the
    single-particle self-energy and vertex corrections. 
    }
    \end{figure}

Turning now to the evaluation of 
$\hat{\chi}^{\rm id}_{\nu,\nu} (\omega, \omega_i)$,
we first note that
the leading high-frequency contribution to 
$\hat{\chi}^{\rm id}_{\nu,\nu} (\omega, \omega_i)$
consists of the two self-energy and two vertex-correction terms
shown in Fig.~9(c).
	The summation of the most singular
diagrams in powers of $(H_1')^2/\omega$ can be  performed by using the 
self-consistent form of the electron-hole propagator \cite{KupcicFA},
as illustrated in Fig.~9(d). 
	(For more details see Ref.~\cite{KupcicFA}.) 
	This approach gives
\begin{eqnarray} 
&& \hat{\chi}^{\rm id}_{\nu,\nu} (\omega, \omega_i)
  \approx  
-\frac{1}{v} \sum_{{\bf k} {\bf k}'\sigma} 
\frac{\partial  f_c({\bf k})}{\partial  E_c({\bf k})}  
 \frac{  \langle | V_1^{cc}({\bf k} - {\bf k}') | ^2 \rangle }{
  \hbar\omega +  \hbar \Sigma^{cc}_{\nu} ({\bf k}, \omega) }
\nonumber \\
&& \hspace{15mm} \times   
\hat{\gamma}^{cc} _{\nu} ({\bf k}, \omega_{i})
\big( \hat{\gamma}^{cc} _{\nu} ({\bf k}, \omega_{i})
 -  \hat{\gamma}^{cc} _{\nu} ({\bf k}', \omega_{i}) \big)
\nonumber \\
&& \hspace{15mm} \times 
\frac{1}{\hbar}  \big[  {\cal D}^{cc}_0 ({\bf k}, {\bf k}', \omega) 
+ {\cal D}^{cc}_0 ({\bf k}', {\bf k}, \omega) \big], 
\label{eq27}
\end{eqnarray}
%
%
where $(1/\hbar) {\cal D}^{cc}_0 ({\bf k}, {\bf k}', \omega)$ 
is a useful abbreviation for
\begin{eqnarray}
\frac{1}{\hbar \omega + E_c({\bf k}) - E_{c}({\bf k}') + {\rm i}\hbar \eta}.
\nonumber
\end{eqnarray}
	Here $\langle \ldots \rangle$ denotes averaging over the
momentum transfer by the disorder.
	$\Sigma^{cc}_{\nu} ({\bf k}, \omega)$ is the channel-dependent 
electron-hole self-energy,
\begin{eqnarray}
 \hbar \Sigma^{cc}_{\nu} ({\bf k}, \omega) 
&=& 
-\sum_{{\bf q}' } \langle | V_1^{cc}({\bf q}' - {\bf k}) | ^2 \rangle
\bigg(  1 -  \frac{\hat{\gamma}^{cc} _{\nu} ({\bf q}', \omega_{i})}{ 
\hat{\gamma}^{cc} _{\nu} ({\bf k}, \omega_{i})} \bigg) 
\nonumber \\  
&&  
\times
\frac{1}{\hbar}\big[  {\cal D}^{cc}_0 ({\bf k}, {\bf q}', \omega) 
+ {\cal D}^{cc}_0 ({\bf q}', {\bf k}, \omega) \big].  
\label{eq28}
\end{eqnarray}
%
%
	The  result of the summation of diagrams in powers of 
$(H_1')^2/\omega$ is thus 
\begin{eqnarray} 
&& \hat{\chi}^{\rm id}_{\nu,\nu} (\omega, \omega_i) 
\nonumber  \\ \nonumber 
&& 
\hspace{3mm} \approx   
\frac{1}{v} \sum_{{\bf k} \sigma} 
\big| \hat{\gamma}^{cc} _{\nu} ({\bf k}, \omega_{i}) \big|^2
\frac{\partial  f_c ({\bf k})}{\partial  E_c({\bf k})}  
\frac{\Sigma^{cc}_{\nu} ({\bf k}, \omega) }{\omega } 
 \\ \nonumber 
&& 
\hspace{8mm} \times
\bigg[1 + \frac{-\Sigma^{cc}_{\nu} ({\bf k}, \omega) }{\omega } 
+ \bigg( \frac{-\Sigma^{cc}_{\nu} ({\bf k}, \omega) }{\omega } \bigg)^2
+ \cdots \bigg] 
\\
&& 
\hspace{3mm} =  
-\frac{1}{v} \sum_{{\bf k} \sigma} 
\big| \hat{\gamma}^{cc} _{\nu} ({\bf k}, \omega_{i}) \big|^2
\frac{\partial  f_c ({\bf k})}{\partial  E_c({\bf k})}  
 \frac{  - \Sigma^{cc}_{\nu} ({\bf k},  \omega) }{
  \omega +   \Sigma^{cc}_{\nu} ({\bf k}, \omega) }.
\label{eq29} 
\end{eqnarray}
%
%
	This  result  has the correct limit for small $\hbar \omega$ 
in comparison with typical damping energies.

It is important to realize here that the expression 
(\ref{eq29}) is obtained under an assumption that is 
valid for the quasi-elastic scattering 
processes, namely that the real part of 
$\Sigma^{cc}_{\nu} ({\bf k}, \omega)$ is negligibly small.
	In this case, we can write 
$\Sigma^{cc}_{\nu} ({\bf k}, \omega) \approx {\rm i}
{\rm Im} \{{\Sigma^{cc}_{\nu}} ({\bf k}, \omega)\}
 \approx {\rm i} \Gamma^{c,{\rm id}}_{\nu}$.
	This can be easily generalized to weakly inelastic 
incoherent scattering by introducing 
$\Gamma^{c,{\rm id}}_{\nu} (\omega)$.
	On the other hand, the introduction of $H_{\rm AF}$, Eq.~(\ref{eq5}),
leads to large coherent effects in 
${\rm Re} \{{\Sigma^{cc}_{\nu}} ({\bf k}, \omega)\}$.
	This  requires the re-examination of the single-particle Hamiltonian 
$H_0 + H_{\rm AF}$, with those coherence effects related to $H_{\rm AF}$
incorporated  also in new effective vertices, and not only in  
${\rm Re} \{\Sigma^{cc}_{\nu} ({\bf k}, \omega)\}$.
	The description of this procedure is postponed to Sec.~VI\,C.

The generalization to other correlation functions gives the universal 
expression 
\begin{eqnarray}
\hat{\chi}_{i ,i}^{\rm id} (\omega) &=& 
\frac{- {\rm i} \Gamma^{c,{\rm id}}_{i}}{\omega + {\rm i} 
\Gamma^{c,{\rm id}}_{i}}
\frac{1}{v_0 } \hat{n}^{\rm id}_{i,i}(\mu),
\label{eq30}
\end{eqnarray}
%
%
for $\Sigma^{cc}_{i} ({\bf k}, \omega) 
\approx {\rm i} \Gamma^{c,{\rm id}}_{i}$, $i = 1, \alpha, \nu$.
	Here $\hat{n}^{\rm id}_{\nu,\nu} (\mu,  \omega_{i})$ 
is the effective channel-dependent density of states at the Fermi energy
of the form
\begin{eqnarray}
\hat{n}^{\rm id}_{\nu, \nu} (\mu,  \omega_{i}) & =& -\frac{1}{N}  
\sum_{{\bf k} \sigma}  
\big| \hat{\gamma}_{\nu  }^{cc} ({\bf k}, \omega_{\rm i}) \big|^2
\frac{\partial f_c ({\bf k})}{\partial E_c ({\bf k})},
\label{eq31}
\end{eqnarray}
%
%
and $\hat{n}^{\rm id}_{1,1} (\mu)$ and 
$\hat{n}^{\rm id}_{\alpha, \alpha} (\mu)$ 
are obtained by  replacing 
$\big| \hat{\gamma}_{\nu  }^{cc} ({\bf k}, \omega_{\rm i}) \big|^2$
in Eq.~(\ref{eq31}) 
with $\big( \hat{q}^{cc} ({\bf k},{\bf k}) \big)^2 = 0$ and 
$\big( \hat{j}^{cc}_{\alpha} ({\bf k}) \big)^2 =
\big( j^{cc}_{\alpha} ({\bf k}) \big)^2$, respectively.
	Also, we define the related effective densities 
$n_{i,j}^{\rm id} (\mu,  \omega_{i})$ and 
$\overline{n}^{\rm id}_{i,j} (\mu,  \omega_{i})$ 
using the total vertices and the constant part of vertices instead of
$\big| \hat{\gamma}_{\nu  }^{cc} ({\bf k}, \omega_{\rm i})\big|^2$
in Eq.~(\ref{eq31}).
	Evidently, $\Gamma^{c,{\rm id}}_1 = 0$ and 
$\hat{\chi}_{1,1}^{\rm id} (\omega) = 0$. 
	Also, 
$n^{\rm d} _{1,1} (\mu) \equiv \hat{n}^{\rm id}_{\alpha, \alpha} (\mu)$
and 
$\Gamma^{c,{\rm d}}_1 \equiv \Gamma^{c,{\rm id}}_{\alpha}$.
 	Both those results are required by the continuity equation 
and the gauge invariance of the intraband optical conductivity
\cite{KupcicFA}.

Let us finally mention the Coulomb screening problem.
	The effects of the Coulomb forces on the indirect processes are
described by the Hopfield series of diagrams shown in Fig.~6(b),
which is an analog of the Hopfield series studied in the context of the 
optical conductivity \cite{Mahan,KupcicFA}.
	This series is free of the $q^{-2}$ singularity
and, for a sufficiently large relaxation rate  $\Gamma^{c,{\rm id}}_{\nu}$ 
(with the critical relaxation rate $\Gamma^{c,{\rm id}}_{\nu,0}$ 
defined precisely in the following subsection), 
does not affect the spectra in a critical manner.
	Therefore, these corrections 
[starting with the second term in Fig.~6(b)] are neglected
in the present analysis, i.e. we take
$\widetilde{\chi}_{\nu, \nu}^{\rm id} (\omega,\omega_i) \approx 
\hat{\chi}_{\nu, \nu}^{\rm id} (\omega, \omega_i)$.

\subsection{Direct vs indirect contributions}
%
%

When the direct and indirect processes are combined, we 
obtain the total Raman correlation function in the form
\begin{equation}
\widetilde{\chi}^{\rm total}_{\nu,\nu}({\bf q},\omega, \omega_{i})  \approx
\widetilde{\chi}^{\rm d}_{\nu,\nu}({\bf q}, \omega, \omega_{i}) 
+ \hat{\chi}^{\rm id}_{\nu,\nu}(\omega, \omega_{i}),
\label{eq32}
\end{equation}
%
%
where
\begin{eqnarray}
&& \widetilde{\chi}^{\rm d}_{\nu,\nu}({\bf q}, \omega, \omega_i) =
\hat{\chi}^{\rm d}_{\nu,\nu} ({\bf q}, \omega, \omega_i)
+
\overline{\chi}^{\rm d}_{\nu,\nu} ({\bf q}, \omega, \omega_i) 
\nonumber \\
&& \hspace{5mm} 
+ \overline{\chi}_{\nu, 1}^{\rm d} ({\bf q}, \omega, \omega_i)
\frac{4 \pi e^2}{ q^2  \varepsilon({\bf q}, \omega)}
\overline{\chi}^{\rm d}_{1, \nu} ({\bf q}, \omega, \omega_i), 
\label{eq33}
\end{eqnarray}
%
%
using again the separation of vertices 
$\gamma^{cc} _{\nu} ({\bf k}, \omega_{i}) = 
\overline{\gamma}^{cc} _{\nu} (\omega_{i}) 
+ \hat{\gamma}^{cc} _{\nu} ({\bf k}, \omega_{i})$
and the corresponding separation of 
$\chi^{\rm d}_{i,j} ({\bf q}, \omega, \omega_i)$.
	There is a well-defined exclusion rule here.
	The constant terms in the vertices participate in the 
direct processes and are screened by the long-range Coulomb forces.
	On the contrary, only the dispersive terms participate in the indirect
processes.
	They are independent of the wave vector {\bf q} and are thus nearly
unaffected by the long-range screening.
	The intensity of the former processes is proportional to small $q^2$,
except in the static metallic limit, 
and the intensity of the latter process  is proportional to 
the channel-dependent relaxation rates $\Gamma^{c,{\rm id}}_{\nu}$.

To find out which of these two processes dominate the correlation
functions of the high-$T_c$ 
cuprates, we now compare the imaginary parts of the expressions 
(\ref{eq25}) and (\ref{eq30}).
	For 
$n^{\rm d}_{\nu,\nu} (\mu, \omega_{i}) 
\approx \hat{n}^{\rm id}_{\nu, \nu} (\mu, \omega_{i})$  and
$\Gamma^{c,{\rm d}}_{\nu} \approx \Gamma^{c,{\rm id}}_{\nu}$,
we obtain the condition $\hbar \omega \approx aq t_{\rm pd}^{\rm eff}$. 
	Furthermore, 
$-{\rm Im} \{ \chi^{\rm id}_{\nu,\nu}(\omega, \omega_{i}) \}$
is characterized by a maximum at $\omega =  \Gamma^{c,{\rm id}}_{\nu}$,
and the critical damping energy is given roughly by 
$\hbar \Gamma^{c,{\rm id}}_{\nu,0} \approx aq t_{\rm pd}^{\rm eff}$, with
$aq \approx 10^{-3}$ typically.
	For the 3D systems and $t_{\rm pd}^{\rm eff} = 1$ eV, the result is 
$\Gamma^{c,{\rm id}}_{\nu,0}/(2 \pi c) \approx 10$ cm$^{-1}$.
	For the usual experimental geometry in the high-$T_c$ cuprates
[$q_{\alpha'} = q_z$ and 
$n^{\rm d}_{\nu,\nu} (\mu, \omega_{i}) \approx (t_{\perp}/t_{\rm pd}^{\rm eff})^2
\hat{n}^{\rm id}_{\nu, \nu} (\mu, \omega_{i})$],  
on the other hand, the critical relaxation rate is 
$\Gamma^{c,{\rm id}}_{\nu,0}/(2 \pi c) \approx aq t_{\perp}$, 
i.e. well below 10 cm$^{-1}$.
	Based on this estimates, for frequencies of the outmost 
experimental interest, $\omega/(2 \pi c) > 50$ cm$^{-1}$,  
the direct processes  can be omitted
and we continue the analysis with the approximate  expression
\begin{equation}
\widetilde{\chi}^{\rm total}_{\nu,\nu}({\bf q},\omega, \omega_{i})  \approx
\hat{\chi}^{\rm id}_{\nu,\nu}(\omega, \omega_{i}).
\label{eq34}
\end{equation}
%
%
	The measured Raman spectra 
$-{\rm Im} \{ \widetilde{\chi}^{\rm total}_{\nu,\nu}(\omega, \omega_{i}) \}$	
are thus proportional to
$-{\rm Im} \{ \hat{\chi}^{\rm id}_{\nu,\nu}(\omega, \omega_{i}) \}$
of Eq.~(\ref{eq30}) for arbitrary $\omega$.

For comparison with experimental and previous theoretical results,
it is useful to rewrite the effective densities 
$\hat{n}^{\rm id}_{\nu, \nu} (\mu, \omega_{i})$ in terms of the related 
densities $n_{i,j}^{\rm id} (\mu, \omega_{i})$, 
which involve the total Raman vertices.
	For this purpose, we notice that the constant terms
$\overline{\gamma}_{\nu  }^{cc} (\omega_{\rm i})$, 
defined by 
$n^{\rm d}_{\nu,1} (\mu, \omega_{i}) = 
\overline{\gamma}_{\nu  }^{cc} (\omega_{\rm i})
n^{\rm d}_{1,1} (\mu)$, 
can be formally expressed
in terms of the effective density of states 
$n_{i,j}^{\rm id} (\mu, \omega_{i})$ 
in the following way
\begin{equation}
\overline{\gamma}_{\nu}^{cc} (\omega_{\rm i}) 
\approx \frac{n^{\rm id}_{\nu,1} (\mu, \omega_{i})}{
n^{\rm id}_{1,1} (\mu)}.
\label{eq35}
\end{equation}
%
%
	This  finally leads to 
\begin{eqnarray}
\hat{n}^{\rm id}_{A_{1g}} (\mu, \omega_{i}) &=& 
\frac{n^{\rm id}_{A_{1g}}(\mu, \omega_{i}) n^{\rm id}_{1,1}(\mu) 
- \big( n^{\rm id}_{A_{1g},1}(\mu, \omega_{i}) \big)^2}{n^{\rm id}_{1,1} (\mu)},
\nonumber \\
\hat{n}^{\rm id}_{\nu} (\mu, \omega_{i}) &=& 
n^{\rm id}_{\nu} (\mu, \omega_{i}), \hspace{3mm} \nu = B_{1g}, \, B_{2g}
\label{eq36}
\end{eqnarray}
%
%
[using the abbreviation 
$\hat{n}^{\rm id}_{\nu} (\mu, \omega_{i}) \equiv 
\hat{n}^{\rm id}_{\nu,\nu} (\mu, \omega_{i})$].

\subsection{Comparison with the  usual  field-theory approach}
%
%

For the sake of comparison with the common  field-theory approaches (FTA)
it is appropriate to notice that
$\widetilde{\chi}^{\rm d}_{\nu,\nu}({\bf q}, \omega, \omega_i)$
of Eq.~(\ref{eq33}) can be rewritten as
\begin{eqnarray}
&& \widetilde{\chi}^{\rm d}_{\nu,\nu}({\bf q}, \omega) =
\bigg[\chi^{\rm d}_{\nu,\nu} ({\bf q}, \omega)
-\frac{\chi^{\rm d}_{\nu,1} ({\bf q}, \omega)
\chi^{\rm d}_{1,\nu} ({\bf q}, \omega)}{
\chi_{1,1} ({\bf q}, \omega)} \bigg]
\nonumber \\
&& \hspace{30mm} 
+ \frac{\chi^{\rm d}_{\nu, 1} ({\bf q}, \omega)
\chi^{\rm d}_{1, \nu} ({\bf q}, \omega)
}{\chi_{1,1} ({\bf q}, \omega)  \varepsilon({\bf q}, \omega)}, 
\label{eq37}
\end{eqnarray}
%
%
in the simplified notation [$\omega_i$ is omitted and it is 
noted that
$\chi^{\rm d}_{1,1} ({\bf q}, \omega) = \chi_{1,1} ({\bf q}, \omega)$].
	The relation (\ref{eq37}) is also the starting point of the  
FTA analyses of the electronic Raman scattering 
\cite{Devereaux2,Monien,Einzel,Sherman2},
and is the source of controversies regarding the role of the long-range
screening  in the Raman scattering.

Most of the  FTA Raman analyses \cite{Zawadowski,Devereaux2,Devereaux5,Sherman2}
use the standard approximation for the
transverse correlation functions \cite{Abrikosov,Mahan}
to study the Raman spectra in the $B_{1g}$ and $B_{2g}$  channels.
	In this case,
$\tilde{\chi}_{\nu,\nu} ({\bf q}, \omega)$ equals 
$\chi_{\nu,\nu} ({\bf q}, \omega)$,
with $\chi_{\nu,\nu} ({\bf q}, \omega)$  given by the  diagram of Fig.~2(a) 
for  $\chi_{\nu,\nu}^{\rm d}  ({\bf q},\omega)$ 
in which the momentum relaxation
is replaced by the energy relaxation.
	Equivalently, this can be formulated by redefining 
the single-electron Green functions  with respect to 
the Green functions used in the charge-charge correlation functions
\cite{Abrikosov,Mahan}.
	For the scattering on the disorder, this leads roughly to 
$\chi_{\nu,\nu} ({\bf q}, \omega) = \chi^{\rm FTA}_{\nu,\nu}(\omega)=
\chi_{\nu,\nu}^{\rm id}  (\omega) - \chi_{\nu,\nu}^{\rm id}  (0)$,
with $\chi_{\nu,\nu}^{\rm id}  (\omega)$ given by Eq.~(\ref{eq30}).
	The same approximation was extended to the $A_{1g}$ channel 
of the high-$T_c$ cuprates in  Ref. \cite{Niksic}. 
	This is a reasonable approximation for the nearly 
half-filled conduction band  with the Raman vertices treated explicitly,
because the resulting ratio $\chi_{A_{1g},1}^{\rm d}  ({\bf q},\omega)/
\chi_{1,1}^{\rm d}  ({\bf q},\omega)$ turns out to be negligibly small, 
as shown below  in Sec.~V\,B.

On the other hand, the usual  approximate description of the Raman
vertices used in the FTA approaches
generates  $\chi_{A_{1g},1}^{\rm d}  ({\bf q},\omega)/
\chi_{1,1}^{\rm d}  ({\bf q},\omega)$  comparable to unity.
	This induces a quite large constant term in the $A_{1g}$
Raman vertex, and, consequently, activates the long-range forces,
as does our approach for a partially filled conduction band. 
	The FTA approaches combine further the Coulomb screening 
in the expression (\ref{eq37}) with the aforementioned approximation for the 
transverse correlation functions 
$\chi_{\nu,\nu} ({\bf q}, \omega)$.
	The Coulomb  term in Eq.~(\ref{eq37}) is first removed 
on taking \cite{Monien,Devereaux2}
$\tilde{\chi}_{1,1} ({\bf q}, \omega) \sim (v_{\rm F} q)^2 /\omega_{\rm pl}^2$, 
i.e. the static screening on the ideal lattice in
$\varepsilon({\bf q}, \omega)$.
	Next, the  momentum relaxation in 
$\chi_{i,j}^{\rm d}  ({\bf q},\omega)$  is replaced 
in the braces of Eq.~(\ref{eq37}) by the energy relaxation.
	Again, this amounts roughly  to the replacement of 
$\chi_{i,j}^{\rm d}  ({\bf q},\omega)$ by 
$\chi^{\rm FTA}_{i,j}(\omega)=
\chi_{i,j}^{\rm id}  (\omega) - \chi_{i,j}^{\rm id}  (0)$.	
In this way, one obtains the common field-theory expression
\cite{Zawadowski,Monien,Devereaux2,Sherman2}
for the screened Raman correlation function in all three channels
$\tilde{\chi}_{\nu,\nu} ({\bf q}, \omega) \approx
\tilde{\chi}^{\rm FTA}_{\nu, \nu}(\omega)$, 
where the $\chi_{i,j}^{\rm d}  ({\bf q},\omega)$ are replaced by 
\begin{eqnarray}
\chi^{\rm FTA}_{i,j}(\omega) 
&=& 
\frac{\omega}{\omega + {\rm i} 
\Gamma^{c,{\rm id}}_{i,j}}
\frac{1}{v_0 } n^{\rm id}_{i,j}(\mu)
\label{eq38}
\end{eqnarray}
%
%
in the braces of Eq.~(\ref{eq37}).
	At frequencies $\omega \ll \omega_{\rm pl}$, the form of the resulting 
${\rm Im} \{ \tilde{\chi}^{\rm FTA}_{\nu,\nu}(\omega) \}$
is thus  quite similar to the imaginary part of our expression 
(\ref{eq34}) [combined with (\ref{eq30}), (\ref{eq35}) and (\ref{eq36})]. 
	The background of this result is that the large Coulomb 
term introduced by FTA for any  band filling is removed therein by the 
static screening.

However, instead of removing the last term in Eq.~(\ref{eq37})
by the use of the static screening on the ideal lattice 
\cite{Monien,Devereaux2},
our approach determines explicitly the role of the long-range forces
in the presence of the disorder for the typical Raman regime
$\omega > v_{\rm F} q_{\alpha}$,
with the Raman vertices treated explicitly.
	It turns out that the last term (negligible for the half-filling) 
is removed from the $A_{1g}$ response
for the partially filled band  by the dynamic, 
rather than by  the static screening of the long-range Coulomb forces
involved in the direct processes.
	This screening is characterized by 
$\chi_{i,j}^{\rm d}  (q_{\alpha},\omega) \propto q_{\alpha}^2 /\omega^2$
in all susceptibilities appearing in the last term of Eq.~(\ref{eq37}).
	In addition, our approach shows immediately that for 
$\omega > \Gamma^{c,{\rm id}}_{i,j}$ Eq.~(\ref{eq37}) is valid in the 
impurity-free form, i.e. that the plasmon peak does not appear in the 
Raman response due to the 
$\chi_{i,j}^{\rm d}  (q_{\alpha},\omega) \propto q_{\alpha}^2 /
\omega^2$ behavior.
	In contrast to that, the FTA does not give a clear recipe how
to extend its treatment of the last term in Eq.~(\ref{eq37})
to the frequencies $\omega \approx \omega_{\rm pl}$.
	It is noteworthy that if   $\varepsilon({\bf q}, \omega)$ 
were to be replaced here by the usual plasma expression 
for the impurity-free lattice but the behavior of other 
$\chi_{i,j}^{\rm d}  ({\bf q},\omega)$'s in this term was kept constant in the 
small {\bf q} limit, using the expression (\ref{eq38}), 
the observation of the plasmon would be predicted in the 
Raman scattering, with a magnitude comparable to that of the 
single-particle term in $\tilde{\chi}^{\rm FTA}_{\nu, \nu}(\omega)$.
	This behavior, common in some semiconductors \cite{Platzman}, 
does not occur in the high-$T_c$ cuprates.

In summary, the Coulomb screening, instead of being all-important 
in the Raman response of the high-$T_c$ cuprates
is not important at any $\omega$.
	Eqs.~(\ref{eq36}) and (\ref{eq30}), although widely used,
are thus derived here for the first time in a consistent manner
for $\omega < \Gamma^{c,{\rm id}}_{i,j}$
and extended to  frequencies around the intraband plasmon frequency.

\section{Intraband Raman spectral functions}
%
%

In order to illustrate the importance of the enhancement of the
electronic Raman spectra by the interband resonance, 
we shall consider now the bare correlation functions
$\chi^{\rm id}_{\nu, \nu} (\omega,\omega_i)$, 
$\nu = A_{1g}, B_{1g}, B_{2g}$, 
in the Drude regime of the $H_{\rm AF} = 0$ case, using 
(i) the static-Raman-vertex approximation, 
$\gamma^{cc} _{\nu} ({\bf k}, \omega_{\rm i}, \omega_{\rm s} )  
\approx \gamma^{cc} _{\nu} ({\bf k})$, usual in most of the current
literature
\cite{Klein,Zawadowski,Monien,Ruvalds,Devereaux1,Devereaux4,Devereaux2,Einzel,Devereaux5}, 
and (ii)  the elastic-Raman-vertex approximation  \cite{Sherman,Sherman2}.
	Also, the reduced correlation function
$\hat{\chi}^{\rm id}_{A_{1g},A_{1g}} (\omega,\omega_i)$ will
be compared to 
$\chi^{\rm id}_{A_{1g},A_{1g}} (\omega,\omega_i)$ 
to estimate the reduction effects present in Eq.~(\ref{eq36}).
	Since, in the numerical calculations discussed  below,
the 3D nature of the problem appears only in the
relaxation rates, which are assumed to be independent of the wave vector
and frequency,  we set $t_{\perp} \approx 0$
and replace the 3D integrations in the correlation functions 
by 2D integrations.

\subsection{Intraband (Drude) Raman scattering}

With $\Sigma^{cc}_{\nu} ({\bf k}, \omega) \approx {\rm i}
\Gamma^{c,{\rm id}}_{\nu}$,
the  spectral functions related to the Drude part of  electronic 
Raman spectra  are given by
\begin{eqnarray}
-{\rm Im} \{ \hat{\chi}^{\rm id}_{\nu,\nu}(\omega, \omega_{i}) \} 
\approx 
\frac{ \omega \Gamma^{c,{\rm id}}_{\nu}} { 
\omega^2 + \big(\Gamma^{c,{\rm id}}_{\nu} \big)^2}
\frac{1}{v_0} \hat{n}^{\rm id}_{\nu} (\mu, \omega_{i}).
\label{eq39}
\end{eqnarray}
%
%
 	For 
$\Gamma^{c,{\rm id}}_{\nu} \approx \Gamma^{c,{\rm id}}$,  the
three Raman channels are still distinguished by the effective 
Raman density of states $\hat{n}^{\rm id}_{\nu} (\mu, \omega_{i})$
(whatever is $t_{pp}$).

Furthermore, the comparison with the  intraband  optical conductivity
\begin{eqnarray}	
{\rm Re} \{ \sigma_{\alpha \alpha}^{c} ( \omega ) \} &=& 
  \frac{\Gamma^{c,{\rm id}}_{\alpha }}{
\omega^2 + \big( \Gamma^{c,{\rm id}}_{\alpha } \big)^2} 
\bigg( \frac{eat_{\rm pd}^{\rm eff} }{\hbar} \bigg)^2 
\frac{1}{v_0}n^{\rm id}_{\alpha} (\mu ),
 \nonumber \\
 \label{eq40}
\end{eqnarray}
%
%
with $(eat_{\rm pd}^{\rm eff} /\hbar)^2 n^{\rm id}_{\alpha} (\mu )/v_0 \equiv 
e^2 n^{\rm eff}_{\alpha \alpha}/m$, where $n^{\rm eff}_{\alpha \alpha}$
is the effective number of conduction electrons per unit cell
(discussed in more detail in Sec.~VI\,A),
gives an analog of the well-known relation valid in  simple Drude metals,
\begin{eqnarray}	
-{\rm Im} \{ \hat{\chi}^{\rm id}_{\nu, \nu} (\omega,\omega_{i}) 
\}
&\propto&
\omega {\rm Re} \{ \sigma_{\alpha \alpha}^{c} ( \omega )\}.
 \label{eq41}
\end{eqnarray}
%
%
	[Notice that 
$\hat{n}^{\rm id}_{\alpha} (\mu ) \equiv n^{\rm id}_{\alpha} (\mu )$,
because the constant term in the current vertex is equal to zero, i.e.
$j_{\alpha} (-{\bf k}) = - j_{\alpha} ({\bf k})$.]
	Here it applies to the CuO$_2$ plane 
($\alpha = x, y$ and $\nu = A_{1g}, B_{1g},  B_{2g}$).
	This relation has been verified in the measured spectra of 
the overdoped high-$T_c$ cuprates
\cite{Uchida,Cooper,Reznik,Sugai,Sugai2,Opel}, 
where the relaxation rates 
$\Gamma^{c,{\rm id}}_{\alpha }$ and $\Gamma^{c,{\rm id}}_{\nu }$ 
have been replaced by 
$\Gamma (\omega) \approx \Gamma (0) 
+ \lambda \omega $.

	The SRVA version of these expressions,
which sets $\omega_{i} = 0$ in Eq.~({\ref{eq39}), was first derived by 
Zawadowski and Cardona \cite{Zawadowski} 
and then extended to the case of strong quasi-particle damping 
($\lambda  \neq 0$)  in Refs.~\cite{Ruvalds,Niksic,Devereaux5}.
	For the overdoped compounds,  
the $\lambda  \neq 0$ single component intraband term is the only 
contribution relevant to the experimental spectra.
	On the other hand, in the underdoped regime, 
the complete model includes 
both the Drude contribution (\ref{eq39}) 
and the contributions of the low-lying  excitations across the AF (pseudo)gap
\cite{Uchida,Cooper,Quijada,Lupi,Venturini},
discussed further in Sec.~VI\,C.

\subsection{Static-Raman vertex approximation}

   \begin{figure}[tb]
    \includegraphics[height=15pc,width=18pc]{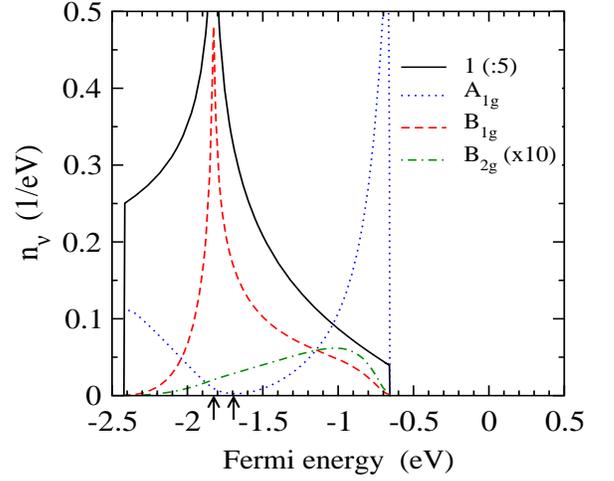}    
    \caption{The dependence of the effective density of states 
    on the Fermi energy $\mu$, for 
    $\Delta_{\rm pd}^{\rm eff} = 0.66$ eV, 
    $t_{\rm pd}^{\rm eff} = 0.73$ eV and $t_{\rm pp} = 0$.  
 	The label $\nu = 1$ denotes the ordinary density of states     
    (divided by a factor of 5),
    and $\nu = A_{1g}$, $B_{1g}$, and $B_{2g}$ correspond to three
    Raman polarizations.
    For clarity the $B_{2g}$ density of states is multiplied by 10.
	The hole picture is used, i.e. the 
    upper band boundary corresponds to the hole doping
    $\delta = 1$ (measured with respect to half-filling).
     The doping range $0 < \delta < 0.3$, relevant to the hole doped
    high-$T_c$ cuprates, is indicated by two arrows.
    }
    \end{figure}

We present now the bare spectra 
$-{\rm Im} \{\chi^{\rm id}_{\nu,\nu}(\omega, \omega_i)\}$
relevant for  the Drude regime.
	First we discuss the validity  of the SRVA
[$\omega_{i} = 0$ in Eq.~({\ref{eq31})].
	As mentioned above, the three-band model  
used in the present calculation  includes  the site energy splitting 
$\Delta_{\rm pd}^{\rm eff} = E_{\rm p} - E_{\rm d}$ 
and the first-neighbor bond-energy $t_{\rm pd}^{\rm eff}$,
but neglects the second-neighbor bond-energy $t_{\rm pp}$
\cite{Emery,Kotliar,KupcicPRB2},
which restricts our physical discussion to the La$_2$CuO$_4$
family, where $t_{\rm pp}$ does not seem to play an all
important  role.

The effective density of states $n_{\alpha} ( \mu )$ 
was evaluated previously \cite{KupcicPC}
for the parameters required to give a reasonable agreement with  the
measured spectral weight of the visible 
conductivity in the La$_2$CuO$_4$-based compounds.
	Using the same parameters, $n_{\nu} ( \mu )$ is
calculated now in  the SRVA.
	Fig.~10 shows this effective density of states, representing 
an appropriate measure for both 
the maxima in the Drude part of the  Raman spectra, Eq.~(\ref{eq39}), 
and  the corresponding spectral weights.
	The most striking result
is that in the  doping range of interest, $0 < \delta < 0.3$,
the ratio $n_{B_{1g}} (\mu) / n_{A_{1g}} (\mu) $ is large
[typically $n_{B_{1g}} (\mu) / n_{A_{1g}} (\mu)  \approx 50$].
	 This enhancement is
related to the fact that, for $t_{\rm pp} = 0$, the factor 
$[\gamma_{A_{1g}}^{cc} ({\bf k})]^2$ becomes negligible 
in comparison with 
$[\gamma_{B_{1g}}^{cc} ({\bf k})]^2$  for the Fermi energy close
to the van Hove energy.	
	This  prediction of SRVA is  however physically unacceptable, 
since the measured
$n_{B_{1g}} (\mu) / n_{A_{1g}} (\mu)  \approx 1$
 \cite{Sugai,Sugai2,Opel}.

Using the definition of the constant terms 
 $\overline{\gamma}_{\nu  }^{cc} (\omega_{\rm i})$, 
$n^{\rm d}_{\nu,1} (\mu, \omega_{i}) = 
\overline{\gamma}_{\nu  }^{cc} (\omega_{\rm i})
n^{\rm d}_{1,1} (\mu)$, we can write 
\begin{eqnarray}
\hspace{5mm}
\hat{n}^{\rm id}_{\nu} (\mu) = n^{\rm id}_{\nu} (\mu) -
\bigg(
\frac{n^{\rm d}_{\nu,1}(\mu)}{n^{\rm d}_{1,1}(\mu)} 
\bigg)^2
n^{\rm id}_{1,1} (\mu)
\label{eq42}
\end{eqnarray}
%
%
in the simplified notation ($\omega_{i}$ is omitted).
	This expression [and its approximate version (\ref{eq36}) as well]
reveals the existence of two qualitatively different regimes:
	(i) For the nearly half-filled conduction band, i.e. for 
the Fermi energy close to the van Hove energy, the second term is 
negligible [$n^{\rm d}_{\nu,1}(\mu)$ crosses zero at $\delta \approx 0.3$,
and $ n^{\rm d}_{1,1}(\mu)$ is singular for 
$\mu \approx \varepsilon_{\rm vH}$]. 	
	For $\mu \approx \varepsilon_{\rm vH}$, the constant term in
the $A_{1g}$ Raman vertex is negligibly small.
	(ii) On the contrary, for the doping well away from  half-filling,
the dispersive terms in the vertices are negligible, leading to the
strong reduction effects in Eq.~(\ref{eq42}) with
$\hat{n}^{\rm id}_{A_{1g}} (\mu) \ll n^{\rm id}_{A_{1g}} (\mu)$.

To simplify the discussion of the resonant effects and the effects of the
AF correlations, in the rest of the article we consider the effective 
density of states
$\hat{n}^{\rm id}_{\nu} (\mu) \approx n^{\rm id}_{\nu} (\mu)$.
	For $0 < \delta < 0.3$, the corrections are of the order 
of few percent, i.e. they are comparable to the effects of the orthorhombic 
distortion on 
$\widetilde{\chi}^{\rm total}_{\nu,\nu}({\bf q},\omega, \omega_{i})$ 
which have been already neglected here.

\subsection{Elastic-Raman vertex approximation}

   \begin{figure}[tb]
    \includegraphics[height=15pc,width=18pc]{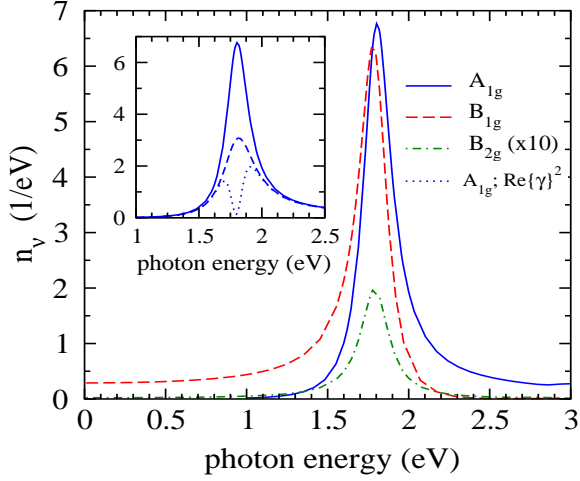}      
    \caption{
    Inset: 
    The resonant enhancement of the
     $A_{1g}$ density of states $n_{A_{1g}}$ for
    $\delta = 0.1$ ($\mu = -1.793$ eV in Fig.~10), and 
    $\hbar \Gamma^{\rm inter} = 0.1$ eV (solid and dotted line) and 0.15 eV 
    (dashed line).
    The dotted (solid, dashed) line represents the contributions of the real
    (real and imaginary) part(s) in 
    $\gamma^{cc} _{\nu} ({\bf k}, \omega_{i} )$ to $n_{A_{1g}}$.
    Main figure: The total bare effective density of states for all three 
    Raman channels for  $\hbar \Gamma^{\rm inter} = 0.1$ eV.
    The $B_{2g}$ spectrum is again multiplied by 10.
    }
   \end{figure}

We calculate therefore the effective density of states 
$n_{\nu} ( \mu, \omega_{i} )$
in the ERVA, i.e. retaining $\omega_{i}$ in Eq.~(\ref{eq31}),
for the hole doping $\delta = 0.1$ and the damping
energies $\hbar \Gamma^{\rm inter} = 0.1$ and 0.15 eV.
	In Fig.~11  we show the results for the model parameters 
used above ($\Delta_{\rm pd}^{\rm eff} = 0.66$ eV and $t_{\rm pd}^{\rm eff} = 0.73$ eV).
	For $\hbar \omega_{i} \approx 0$, the large 
$n_{B_{1g}} (\mu, \omega_{i})$ intraband term, associated with van Hove 
singularities, is large with respect to the interband 
$n_{A_{1g}} (\mu, \omega_{i})$ term.
	For $\hbar \omega_{i}$ around  $E_N({\bf k})- \mu \approx$ 1.8 eV
($N$ for the non-bonding band), 
the resonant (interband) contribution to $n_{A_{1g}} (\mu, \omega_{i})$
is nearly equal to the sum of the static (intraband) and resonant (interband)
terms in $n_{B_{1g}} (\mu, \omega_{i})$.
	In the maximum, the comparable interband contributions dominate.
	This energy range corresponds to 
$E_N({\bf k})- \mu < \hbar \omega_{i} 
< (E_N({\bf k})- E_c({\bf k}))_{\rm max}$,
because for $t_{\rm pp}=0$, the optical excitations between the
conduction and antibonding bands are negligible \cite{KupcicPRB2}.
	For $t_{\rm pp}$ large enough, the latter excitations become important 
as well, and resonant effects are extended to the energy region
$E_N({\bf k})- \mu < \hbar \omega_{i} < (E_P({\bf k})- E_c({\bf k}))_{\rm max}$
(i.e. between 1.7 and 4 eV).
	Due to the resonant enhancement of the Raman scattering processes, 
we find the ratio 
$ n_{B_{1g}} (\mu, \omega_{i}) /n_{A_{1g}} (\mu, \omega_{i})  $ 
consistent with the experimental observation.
	Notice, however, the reduction of the resonant effect 
with increasing damping energy $\hbar \Gamma^{\rm inter}$ 
(inset of the figure).

The spectral weight  of the $B_{2g}$ channel relative to two other channels 
turns out to be  one order of magnitude
smaller than the one usually found in experiments.
	This reflects the fact that various processes described by 
other parameters of the three-band model, and in particular by
the direct oxygen-oxygen hopping  $t_{\rm pp}$,  are absent here.
  	It should be noticed that  $t_{\rm pp}$ opens
an additional channel in the electron-photon coupling [see Eq.~(\ref{eqA3})]
involving predominantly the electronic states in the nodal $k_x=k_y$ 
region of the Fermi surface.
	As easily seen, this leads in the first place to the enhancement 
of the $B_{2g}$ Raman spectra giving the contributions proportional  
to $t_{\rm pp}$ in $\gamma_{B_{2g}}^{cc} ({\bf k}, \omega_i)$,
additional to the contributions of the indirect oxygen-oxygen hopping 
processes [$\propto (t_{\rm pd}^{\rm eff})^2$] shown in Figs.~10--11.

	We notice finally that, if the  contributions 
of Im$\{ \gamma^{CC}_{\nu} ({\bf k}, \omega_{i} ) \}$ to 
$n_{\nu} (\mu, \omega_{i})$ are neglected, 
one obtains the resonant structure characterized by
two peaks split  approximately by  the energy $2 \hbar \Gamma^{\rm inter}$, 
as represented 
in the inset of Fig.~11 by the dotted line.
	Similar  dependence of the Raman spectra on the photon frequencies
was already proposed in the multiband study  of the electron-mediated 
photon-phonon coupling functions \cite{Sherman}.

It should be noticed that most of the recent Raman studies are focussed	
only on the $B_{1g}$ and $B_{2g}$  channels.
	These two channels scan the complementary parts of the Fermi
surface (the vicinity of the van Hove points in  $B_{1g}$  
and the nodal region of the Brillouin zone in $B_{2g}$) and 
almost all relevant physics is present
in the related spectra \cite{Opel,Naeini,Venturini}.
	Our comparison with the experimental data, given in Sec.~VI\,C, will 
be thus also limited to these two channels.

\section{Effects of the AF ordering}
%
%

In order to make our analysis of the coherence factors analytically 
tractable we shall restrict it here to the situations in which the direct
oxygen-oxygen hopping $t_{\rm pp}$ is not qualitatively important
and set it equal to zero.
	Such is the case of La$_{2}$CuO$_4$ based families 
for the doping not too 
far from the optimal doping, where the Fermi surface is nearly  square.

The effective AF potential $\Delta ({\bf k})$ is
assumed to be of  the $d_{x^2-y^2}$  symmetry,
$\Delta ({\bf k})  = 0.5 \Delta_{\rm AF}
(\cos {\bf k} \cdot {\bf a}_1 - \cos {\bf k} \cdot {\bf a}_2)$.
	This  potential dominantly affects the states
close to the van Hove points, leads to the dimerization of the 
bands, and  is accompanied by the low-lying interband  processes
characterized by a threshold energy proportional to the magnitude 
$\Delta_{\rm AF}$.
	Two subbands of the conduction band will be denoted by the indices 
$L = C$ (upper band) and $L = \underline{C}$ (lower band).
	For the half-filled conduction band of the $t_{\rm pp} = 0$ model,
${\bf Q}_{\rm AF}$ leads to the ideal nesting of the Fermi surface and, 
correspondingly, the relevance of this perturbation grows with 
decreasing hole doping.

In other cases, the interplay between $t_{\rm pp}$ and
$\Delta ({\bf k})$  is probably 
responsible for the anomalies regarding e.g. the development of both the 
Fermi surface shape \cite{DKS}
and the optical conductivity with doping.
	Namely, for $t_{\rm pp}$ large enough with respect to 
$t_{\rm pd}^{\rm eff}$, even small changes in the 
hole doping could produce dramatic changes in the electrodynamic features
of the electron system (this might be
analogous to the situation found in the quasi-one-dimensional
Bechgaard salts 
\cite{Degiorgi}).
	ARPES measurements in the YBa$_2$C$_3$O$_{7-x}$ and Bi-based 
cuprates \cite{Norman,Shabel}
are indicative of such a regime, not discussed here.

\subsection{Hall coefficient }
%
%

In the three-band model with the magnetic field normal 
to the conduction
plane, the room-temperature Hall coefficient  is of the form
$R_{\rm H} \approx 1/(ecn_{\rm H})$, where $n_{\rm H}$ 
is the effective Hall number
given by $n_{\rm H} = n_{xx}^{\rm eff} n_{yy}^{\rm eff} / n_{xy}^{\rm eff}$.
	The diagonal and off-diagonal effective numbers of
charge carriers read as \cite{Ziman,Kontani,KupcicPC}
\begin{eqnarray}
n_{\alpha \alpha}^{\rm eff}  &=&
-\frac{m}{e^2} \frac{1}{v} \sum_{{\bf k}^* \sigma}
\big[ J_{ \alpha}^{CC} ({\bf k}) \big]^2
\frac{\partial f_{C} ({\bf k})}{\partial E_C ({\bf k})},
\label{eq43} \\
n_{xy}^{\rm eff} &=& \frac{m}{e^2}\frac{1}{v} \sum_{{\bf k}^* \sigma}
\frac{\partial f_{C} ({\bf k})}{\partial E_C ({\bf k})}
J_x^{CC} ({\bf k})[ \gamma_{yy}^{CC} ({\bf k}) J_x^{CC} ({\bf k})
 \nonumber \\ &&
\hspace{15mm} -\gamma_{xy}^{CC} ({\bf k}) J_y^{CC} ({\bf k})]
\label{eq44}
\end{eqnarray}
%
%
($\alpha = x$ or $y$).
	The structure of the intraband current vertices, 
$J_{\alpha}^{CC} ({\bf k})$ [$J_{\alpha}^{cc} ({\bf k})$], 
and the static Raman vertices, 
$\gamma_{\alpha \beta}^{CC} ({\bf k})$
[$\gamma_{\alpha \beta}^{cc} ({\bf k})$], for the 
$\Delta_{\rm AF} \neq 0$ ($\Delta_{\rm AF} = 0$) case is determined in Appendix  B (A).
	For $\Delta_{\rm AF} \neq 0$ ($\Delta_{\rm AF} = 0$),
\hbox{${\bf k}^*$ (${\bf k}$)} refers to the new (old) Brillouin zone.	
	The DC conductivity can be scaled by the diagonal effective
numbers, as well,  according to the relations (\ref{eq40}) and (\ref{eq43}),
$\sigma_{\alpha \alpha}^{\rm DC} =  
e^2 n_{\alpha \alpha}^{\rm eff}/(m \Gamma^{c,{\rm id}}_{\alpha})$.

   \begin{figure}[t]
    \includegraphics[height=15pc,width=18pc]{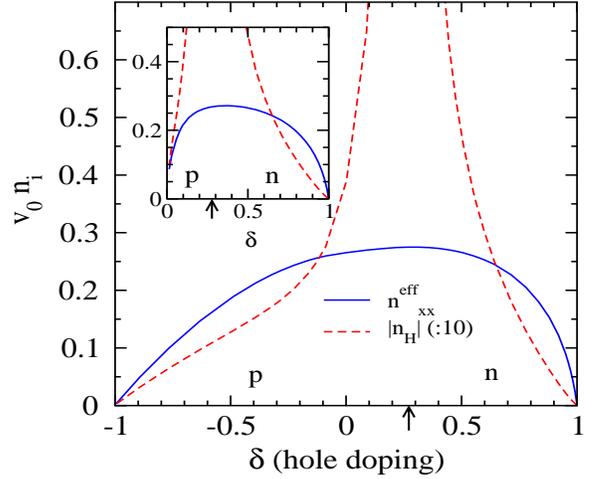}    
    \caption{
    Main frame: Effective numbers $n_{xx}^{\rm eff}$ and
    $|n_{\rm H}|$ (representing also 
    the DC conductivity and the inverse Hall coefficient, 
    scaled by $ e^2/(mv_{0}\Gamma^{c,{\rm id}}_{x})$ and $ec/v_{0}$,
     respectively)
    as a function of the doping level for $\Delta_{\rm AF}  = 0$.
    Inset: The effect of the AF correlations on 
    $n_{xx}^{\rm eff}$ and $|n_{\rm H}|$  for the $d_{x^2-y^2}$ symmetry
    perturbation $\Delta ({\bf k})$ with  $\Delta_{\rm AF} = 50$ meV,
    in the hole-doped region.	
    	The critical doping $\delta_0$, where 
    $n_{xy}^{\rm eff} = 0$, is labeled by arrows.
    	$n$ and $p$ denote, respectively, the region of electron-like    
    ($n_{\rm H}, n_{xy}^{\rm eff}  < 0$) and hole-like 
    ($n_{\rm H}, n_{xy}^{\rm eff}  > 0$) behavior of the charge carriers.
    }
    \end{figure}

The effective numbers (\ref{eq43}) and (\ref{eq44}) 
are  extremely sensitive to the  correlation effects.
	In order to illustrate this dependence, 
the numbers  $n_{xx}^{\rm eff}$ and $n_{\rm H}$	are  calculated  
with  and without  the potential $\Delta ({\bf k})$ of the $d_{x^2-y^2}$ 
symmetry and are compared to the experimental observations in
La$_{2-x}$Sr$_x$CuO$_4$ \cite{Uchida,Uchida2}
showing that (i) the change of the sign of $n_{\rm H}$
occurs nearly at $\delta_0 \approx 0.25$; 
(ii) $n_{\rm H} \propto \delta$ in the underdoped compounds; and
(iii) $n_{\alpha \alpha}^{\rm eff} \propto \delta$ for
$\delta \rightarrow 0$.
	The results are given in Fig.~12 for
$\Delta_{\rm AF}  = 0$ and 50 meV.
	The main figure illustrates the well-known fact that for a pair of 
bonding and antibonding bands the critical doping $\delta_0$,
which separates the electron-like doping region(s) from the 
hole-like one(s), is shifted for finite 
$t_{\rm pd}^{\rm eff}/\Delta_{\rm pd}^{\rm eff}$ 
($t_{\rm pp}=0$)
from $\delta = 0$ in the positive (negative)
direction for the lower (upper) band,
breaking in this way a simple electron-hole symmetry in each of these
two bands.
	For the wide conduction band, characterized by 
$\Delta_{\rm pd}^{\rm eff} = 0.66$ eV and $t_{\rm pd}^{\rm eff} = 0.73$ eV, 
this results in $\delta_0 \approx 0.27$, in agreement with the observation (i).
	The measured linear $\delta$-dependences of 
$n_{xx}^{\rm eff}$ (iii) and $n_{\rm H}$ (ii) 
can be related to the mid-infrared (MIR) gap structure,
as  seen from the inset of Fig.~12.
	It should also be noticed that, for $\Delta_{\rm AF}$ not too large,
the position of $\delta_0$ is only slightly dependent on $\Delta_{\rm AF}$.
	More importantly, due to the doubled number of zeros of
$\partial^2 E_C ({\bf k}) /\partial k_{\alpha} \partial k_{\beta}$
(which appear above and below the original van Hove energy 
$\varepsilon_{\rm vH}$), the effective number 
$n_{xy}^{\rm eff}$ has two zeros, resulting in an 
additional critical doping within  the electron-doped range.
	In  crude terms, this restores the electron-hole symmetry 
of the phase diagram of the high-$T_c$ cuprates, 
which is   seen in the Hall coefficient 
measurements \cite{Uchida2,Uchida}.

\subsection{Optical conductivity }
%
%

The dependence of the low-frequency  conductivity 
on the symmetry and magnitude of the dimerization potential  
$\Delta ({\bf k})$ is analyzed in detail in 
Refs.~\cite{KupcicPC,KupcicFA}.
	For the sake of completeness we enumerate here 
the most important results.
	The two-component $\Delta_{\rm AF} \neq 0$ intraband conductivity reads 
\begin{eqnarray}
 \sigma^{\rm intra}_{\alpha \alpha} ( \omega) &\approx& 
 \zeta_1 \frac{{\rm i}}{\omega} 
\frac{e^2 n_{\alpha \alpha}^{\rm eff}}{m}
\frac{ \omega}{ \omega + {\rm i} \Gamma^{c,{\rm id}}_{\alpha}} 
-\zeta_2 {\rm i}  \omega \alpha^{\rm MIR}_{\alpha \alpha} (\omega), 
\nonumber \\
\label{eq45}
\end{eqnarray}
%
%
with the effective number of conduction electrons, 
$ n_{\alpha \alpha}^{\rm eff}$, and the MIR polarizability, 
$\alpha^{\rm MIR}_{\alpha \alpha} (\omega)$, given by 
\begin{eqnarray}
n_{\alpha \alpha}^{\rm eff} &=& \frac{1}{v} \sum_{{\bf k}^* \sigma} 
\gamma^{CC}_{\alpha \alpha} ({\bf k})  [1 -  f_C({\bf k})],
\label{eq46} \\
\alpha^{\rm MIR}_{\alpha \alpha} (\omega) &=& 
\frac{1}{ \omega^2} \frac{1}{v} \sum_{{\bf k}^*  \sigma }  
\frac{(\hbar \omega)^2 |J_{\alpha}^{C \underline{C}} ({\bf k})|^2}{
E^2_{C \underline{C}} ({\bf k}) }
\nonumber \\ 
&& \times
\frac{ 2E_{C \underline{C}} ({\bf k})[f_C({\bf k}) -1]  }{
(\hbar \omega  + {\rm i} \hbar \Gamma^{\rm MIR}_{ \alpha} )^2 
- E^2_{C \underline{C}} ({\bf k})  }. 
\label{eq47}
\end{eqnarray}
%
%
	The  renormalization factors $\zeta_1$ and $\zeta_2$ 
in Eq.~(\ref{eq45}) 
serve here to model the effects of fluctuations of  auxiliary bosons
on the low-frequency optical excitations \cite{KupcicPC}.
	The vertex $J_{\alpha}^{C \underline{C}} ({\bf k})$ 
and the energy difference $E_{C \underline{C}} ({\bf k})$
are  given in Appendix B.

   \begin{figure}[t]
    \includegraphics[height=15pc,width=18pc]{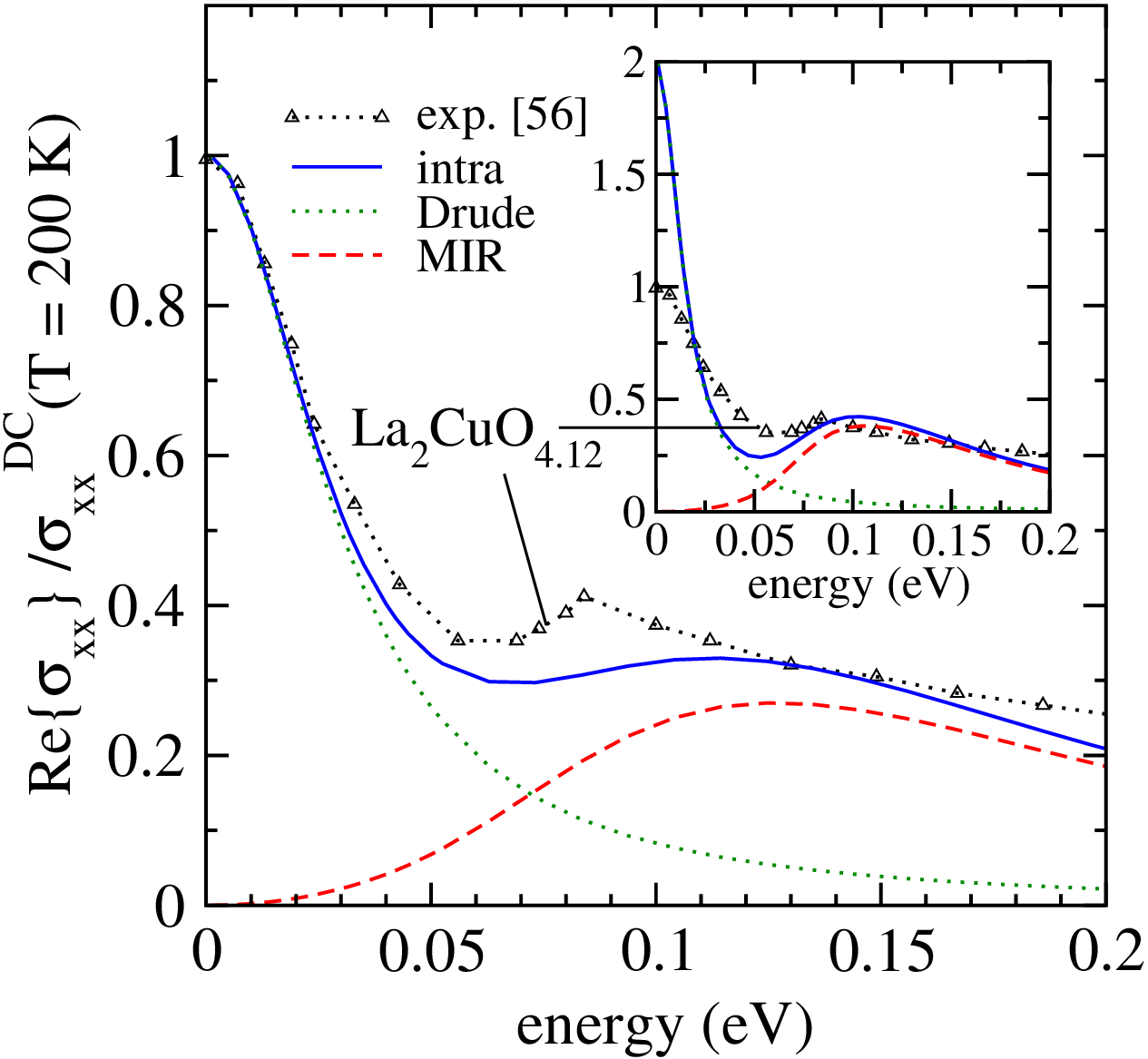}    
    \caption{  
    The  optical conductivity (\ref{eq45})   
    for the anisotropic-s potential 
    $\Delta ({\bf k})  = \Delta_{\rm AF} [0.5 +0.125
(\cos {\bf k} \cdot {\bf a}_1 - \cos {\bf k} \cdot {\bf a}_2)^2]^{1/2}$
with  
    $\Delta_{\rm AF} = 45$ meV,
    $\Delta_{\rm pd}^{\rm eff} = 0.66$ eV, $t_{\rm pd}^{\rm eff} = 0.73$ eV, 
    $\delta = 0.1$,
    $\zeta_1 = 0.18$ and $\zeta_2 = 0.4$.    
    Main figure: $\hbar \Gamma^{c,{\rm id}}_{\alpha} = 30$ meV and 
    $\hbar \Gamma^{\rm MIR}_{\alpha} = 50$ meV (suitable to $T = 200$ K spectra
    in the La$_2$CuO$_4$ based compounds).
    Inset: $\hbar \Gamma^{c,{\rm id}}_{\alpha} = 15$ meV and 
    $\hbar \Gamma^{\rm MIR}_{\alpha} = 25$ meV ($T \approx 100$ K).
    The data measured in La$_{2}$CuO$_{4.12}$ at $T = 200$ K 
    \cite{Quijada} connected by the dotted line
    are given for comparison.    	
    }
    \end{figure}

Fig.~13 illustrates the typical low-frequency  spectra measured 
in La$_2$CuO$_{4.12}$, compared to the model predictions. 
	In spite of its simplicity, the model (\ref{eq45})--(\ref{eq47})
with $t_{\rm pp} = 0$ can explain why the MIR structure 
in La$_{2}$CuO$_{4.12}$ 
is nearly independent of temperature \cite{Quijada}.
	Namely, at temperatures below the room temperature, 
the position of the MIR maximum $\hbar \omega_{\rm MIR} \approx 90$ meV is 
well above the relaxation rate $\hbar \Gamma^{\rm MIR}_{\alpha}$
and correspondingly $\hbar \omega_{\rm MIR} \approx  2 \Delta_{\rm AF}$,
independent of $\hbar \Gamma^{\rm MIR}_{\alpha}$.
	This situation strongly contrasts with those observed in the 
Bechgaard salts  \cite{Degiorgi}
or in Bi$_2$SrCuO$_6$   \cite{Lupi}
where small Drude spectral weights
(i.e. $v_0 n^{\rm eff}_{\alpha \alpha} \ll 1$) reveal the
interplay between $t_{\rm pp}$  (or $t_{\rm b}$ in the Bechgaard salts) 
and the energy scale $ 2 \Delta_{\rm AF}$
\cite{KupcicPB}.

\subsection{$B_{1g}$ and $B_{2g}$ Raman spectra}
%
%

   \begin{figure}[tb]
    \includegraphics[height=20pc,width=18pc]{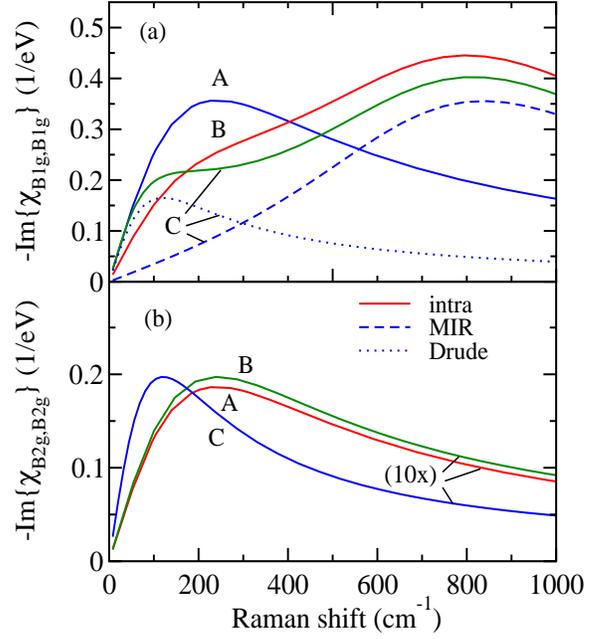}    
    \caption{The   $B_{1g}$  (a) and $B_{2g}$ (b)
    electronic Raman spectra obtained by ERVA for the $d_{x^2-y^2}$  symmetry
    potential $\Delta ({\bf k}) $.
    The parameters are     
    $\Delta_{\rm pd}^{\rm eff} = 0.66$ eV, $t_{\rm pd}^{\rm eff} = 0.73$ eV, 
    $\delta = 0.1$,
    $\hbar \omega_{i} = 2$ eV, $ \hbar \Gamma^{\rm MIR}_{\nu} = $ 50 meV and
    $ \hbar \Gamma^{\rm inter} = $ 0.1 eV.
     The curves A (B): $\Delta_{\rm AF}  = 0 $ (45) meV 
     and $\hbar \Gamma^{c,{\rm id}}_{\nu} = $ 30 meV.
     The curves C: $\Delta_{0} = 45$ meV  
     and $\hbar \Gamma^{c,{\rm id}}_{\nu} = $ 15 meV
     [with the Drude (dotted line) and MIR (dashed line) contributions
     indicated as well].  
    The $B_{2g}$ spectrum is multiplied by 10.   
   }
    \end{figure}

Next, we extend the discussion of the AF   effects 
to the electronic Raman spectra.
	In the hole-doped regime, the Drude-like contributions 
and the low-lying transitions through the AF (pseudo)gap
are given by Eq. (\ref{eq39}) and by
\begin{eqnarray}
&&-{\rm Im} \{ \chi ^{{\rm MIR}}_{\nu, \nu}(\omega, \omega_1)\}  \approx 	
\frac{1}{N}  \sum_{{\bf k}^* \sigma}  
 |\gamma_{\nu  }^{C\underline{C}} ({\bf k}, \omega_1)|^2[f_C({\bf k}) -1]
\nonumber  \\ 
&& \hspace{20mm} \times
{\rm Im} \bigg\{ \frac{ 2E_{C \underline{C}} ({\bf k})  }{
(\hbar \omega  + {\rm i} \hbar \Gamma^{\rm MIR}_{\nu} )^2 
- E^2_{C \underline{C}} ({\bf k})  }  \bigg\} ,
 \label{eq48}
\end{eqnarray}
%
%
respectively.
	Neglecting the effects of $\Delta ({\bf k})$ on the 
intermediate interband processes and applying the static approximation 
for the low-frequency part of
the Raman vertex, the elastic Raman vertices in the expressions 
(\ref{eq39}) and (\ref{eq48}) calculated at $t_{\rm pp} = 0$
are given by 
\begin{eqnarray}
&&\gamma_{\nu  }^{CC} ({\bf k}, \omega_i) \approx 
\gamma_{\nu  }^{cc} ({\bf k}, \omega_i) 
\cos ^2 \frac{\varphi ({\bf k})}{2} 
\nonumber \\ 
&& \hspace{25mm} + \gamma_{\nu  }^{cc} ({\bf k}\pm {\bf Q}_{\rm AF}, \omega_i) 
\sin ^2 \frac{\varphi ({\bf k})}{2} ,
\nonumber  \\ 
&&\gamma_{\nu  }^{C\underline{C}} ({\bf k}, \omega_i) \approx 
\frac{1}{2}[\gamma_{\nu  }^{cc} ({\bf k}, \omega_i)
- \gamma_{\nu  }^{cc} ({\bf k}\pm {\bf Q}_{\rm AF}, \omega_i) ]
\sin \varphi ({\bf k}) .
\nonumber \\ 
\label{eq49}
\end{eqnarray}
%
%
	$\varphi ({\bf k})$ is an auxiliary phase defined in Appendix B.

Again,  for $\Delta_{\rm AF} \ll t_{\rm pd}^{\rm eff}, \Delta_{\rm pd}^{\rm eff}$, 
the $d_{x^2-y^2}$ symmetry   of $\Delta ({\bf k})$ 
causes significant effects 
in the Raman spectra only for relatively small doping ($ \delta < 0.15$)
when the Fermi energy $\mu$ is close to the van Hove singularities.
	The most important qualitative results are illustrated in Fig.~14
for  $\delta = 0.1$ and $\Delta_{\rm AF} =$ 0, 45 meV.

First of all, we observe in Fig.~14 that the MIR peak 
in the optical conductivity  is accompanied by a similar peak in the
Raman spectra, but only in the $B_{1g}$ channel.
	As a result, the Raman spectral density increases with frequency 
towards a maximum in the $B_{1g}$ channel, at 
$\hbar \omega \approx 2\Delta_{\rm AF}$, in contrast to the 
$B_{2g}$ channel, where it decreases immediately after the frequency 
$\hbar \omega \approx \hbar \Gamma^{c,{\rm id}}_{\nu}$.
This agrees qualitatively with the Raman experimental results.

Second, the observed \cite{Naeini}
doping-induced weakening of the Drude part of the  $B_{1g}$ spectra by
one order of magnitude with respect to the $B_{2g}$ spectra 
below $\delta \approx 0.15$, 
can be related to the (pseudo)gap
features in the electron dispersion in the vicinity of the original 
van Hove points.
	Namely,  the $B_{1g}$ effective 
density of states at the Fermi level (shown in Fig.~10) 
is strongly suppressed  for
$|\varepsilon_{\rm vH} - \mu| < \Delta_{\rm AF}$. 
	This contrasts with the 
$B_{2g}$ case where the spectra come dominantly from the nodal region of the 
Fermi surface, unaffected by $\Delta ({\bf k})$.

\section{Conclusion}

The electronic Raman correlation functions have been calculated here for 
the Emery three-band model,  using the distinction between the 
direct and indirect scattering on the quasi-static disorder.  
	It is shown that there is a simple exclusion rule
connecting these two scatterings and the long-range Coulomb screening.
	The direct processes concern the constant terms in the vertices.
	They are strongly affected by the long-range screening,
and, in the dynamic limit, participate in the correlation functions
through the contributions proportional to small $q^2$.
	The indirect processes include only the dispersive terms in the 
vertices.
	They are nearly unaffected by the long-range forces, and their
contributions to the correlation functions are proportional to 
the channel-dependent relaxation rates.
	It is shown so that in the high-$T_c$ cuprates the contributions
of the direct processes to the Raman correlation functions can be safely
neglected.
	Using the elastic approximation for the Raman vertices
in two [with and without the AF dimerization gap 
$\Delta ({\bf k})$] analytically 
solvable versions of the $t_{\rm pp} = 0$ Emery three-band model,
we show than that the resonant Raman scattering processes remove a large 
discrepancy between the spectral weights of the $A_{1g}$ and $B_{1g}$
Raman channels obtained in the static approximation for the Raman vertices.
	The resulting spectra  agree reasonably well with experimental findings.
	It is also shown that
the anomalous MIR peak in the optical conductivity, 
observed  in the underdoped compounds, is correlated
with the corresponding structure which appears only 
in the $B_{1g}$ Raman channel,
as well as with the measured linear $\delta$-dependence 
of the Hall number. 
	This relation is explained here in terms of the 
$\Delta ({\bf k}) \neq 0$ AF correlations.
	On the other hand, the $\Delta ({\bf k})= 0$ Emery model used
to fit the overall band structure, a part of which is seen 
in the  ARPES data \cite{IMR}, leads to different results.
	Particularly important in this respect are  
Raman selection rules.
	The small energy scales observed in the Raman scattering,
just as in the ARPES data \cite{DKS}, are therefore better related 
to the AF correlations within
the conduction band than to the low-energy interband transitions in 
the strongly correlated  $\Delta ({\bf k})= 0$ metallic state.

\section*{Acknowledgement}

We acknowledge useful correspondence with Prof. A. Zawadowski.
	This research was supported by the Croatian Ministry of Science 
and Technology under Project \hbox{0119-256}.

\appendix

\section{Three-band vertex functions}

The coupling of the vector potential ${\bf A} ({\bf r})$
to the conduction electrons of the 
Emery three-band model is given in the usual way
\cite{KupcicPB}, 
by replacing the hole creation
(and annihilation) operators in the bare Hamiltonian $H_0$
by
\begin{eqnarray} 
\tilde{l}^{\dagger}_{n \sigma} &=&  l^{\dagger}_{n \sigma}
e^{\mathrm{i}e  /(\hbar c) ({\bf R}_n + {\bf r}_l) \cdot  
{\bf  A} ( {\bf R}_n + {\bf r}_l )}
\label{eqA1}
\end{eqnarray}
%
%
 (similar for $\tilde{l}_{n \sigma}$).
 	Here ${\bf R}_n$ and ${\bf r}_l$ are, respectively, 
the Bravais lattice vector and the position in the primitive cell 
of the orbital labeled by the index $l$.
	The Taylor expansion in the vector potential of 
$\widetilde{H}_0$ to the second order leads to 
\begin{eqnarray}
\widetilde{H}_0 - H_0 &\approx &  H^{\rm ext} =
\sum_{ll'{\bf k}{\bf q} \sigma}  \delta H_0^{ll'} ({\bf k},{\bf q}) 
l_{{\bf k} + {\bf q} \sigma}^{\dagger} l'_{{\bf k} \sigma},
\label{eqA2}
\end{eqnarray}
%
%
where 
\begin{eqnarray}
&&\! \! \! \!  \! \! \! \! \! \! \! \!  \! \! \! \! 
\delta  H_0^{ll'} ({\bf k},{\bf q})  \approx    
- \frac{1}{c}
\frac{e}{\hbar } \sum_{\alpha} 
\frac{\partial  H_0^{ll'} ({\bf k})}{\partial k_{\alpha}}
 A_{\alpha} ({\bf q}) 
\nonumber \\   
 &&
+ \frac{e^2}{2mc^2} 
\frac{m}{\hbar ^2}
\sum_{{\bf q}' \alpha \beta}
\frac{\partial^2 H_0^{ll'} ({\bf k})}{\partial k_{\alpha} \partial k_{\beta}}
A_{\alpha}  ({\bf q}-{\bf q}') A_{\beta}  ({\bf q}'). 
\label{eqA3} 
\end{eqnarray}
%
%
	In the Bloch representation,  $H^{\rm ext}$ is given by 
the expression (\ref{eq9}), with the vertex functions
\begin{eqnarray}
&& J_{\alpha}^{LL'} ({\bf k}) = 
\frac{e}{\hbar } \sum_{ll'} 
\frac{\partial  H_0^{ll'} ({\bf k})}{\partial k_{\alpha}}
U_{{\bf k}} (l,L) U^*_{{\bf k}} (l',L'),
\nonumber  \\ 
&& \gamma ^{LL'}_{\alpha \beta} ( {\bf k};2)  = 
-\frac{m}{\hbar ^2} \sum_{ll'}
\frac{\partial^2 H_0^{ll'} ({\bf k})}{\partial k_{\alpha} \partial k_{\beta}}
U_{{\bf k}} (l,L) U^*_{{\bf k}} (l',L')
\nonumber \\ 
\label{eqA4} 
\end{eqnarray}
%
%
($\alpha, \beta = x$, $y$).

The number of channels in the electron-photon coupling is equal to the number 
of independent bond energies; $t_{\rm pd}^{\rm eff}$ 
and $t_{\rm pp}$ in the Emery 
three-band model for the in-plane processes.
	For  the $t_{\rm pp} = 0$ three-band model, 
one obtains  the dimensionless in-plane current 
and bare Raman vertices ($\alpha = x$ or $ y$) of the form
\cite{KupcicPC}
\begin{eqnarray}
j^{cc}_{\alpha} ({\bf k}) &=& 
t_{\rm pd}^{\rm eff} \frac{2 u_{\bf k} v_{\bf k}}{t_{\bf k}} 
\sin {\bf k} \cdot {\bf a}_{\alpha},
\nonumber  \\
j^{cP}_{\alpha} ({\bf k}) &=& 
  t_{\rm pd}^{\rm eff}\frac{u^2_{\bf k} -v^2_{\bf k}}{t_{\bf k}} 
\sin {\bf k} \cdot {\bf a}_{\alpha},
\nonumber \\
j^{cN}_{x} ({\bf k}) &=& 
t_{\rm pd}^{\rm eff} \frac{2u_{\bf k} }{t_{\bf k}} 
\sin \frac{1}{2}{\bf k} \cdot {\bf a}_{2}
\cos \frac{1}{2}{\bf k} \cdot {\bf a}_{1},
\nonumber \\
j^{cN}_{y} ({\bf k}) &=&
- t_{\rm pd}^{\rm eff}\frac{2u_{\bf k} }{t_{\bf k}} 
 \sin \frac{1}{2}{\bf k} \cdot {\bf a}_{1}
\cos \frac{1}{2}{\bf k} \cdot {\bf a}_{2},
\label{eqA5}
\end{eqnarray}
%
%
and
\begin{eqnarray}
\gamma ^{cc}_{\alpha \beta} ({\bf k};2) &=& 
\delta_{\alpha, \beta}\frac{m}{m_{xx}}
\frac{\Delta_{\rm pd}^{\rm eff} u_{\bf k} v_{\bf k}}{t_{\bf k}} 
\sin ^2 \frac{1}{2}{\bf k} \cdot {\bf a}_{\alpha},
\label{eqA6}
\end{eqnarray}
respectively, with
\begin{eqnarray}
J^{LL'}_{\alpha} ({\bf k}) &=& 
\frac{e a t_{\rm pd}^{\rm eff}}{\hbar } j^{LL'}_{\alpha} ({\bf k}). 
\label{eqA7}
\end{eqnarray}
%
%
	$u_{\bf k}$,  $v_{\bf k}$, and $t_{\bf k}$ are the auxiliary functions
defined in Ref. \cite{KupcicPRB2}, and 
$m_{xx} = \hbar^2 \Delta_{\rm pd}^{\rm eff} /( 2a^2 (t_{\rm pd}^{\rm eff})^2)$ 
is the in-plane mass scale ($|{\bf a}_1| =|{\bf a}_2| = a$).

\section{Vertex functions with AF}
The AF dimerization of the conduction band $E_c ({\bf k})$  
caused by $H_{\rm AF}$ is solved elsewhere \cite{KupcicPC}.
	Apparently, $H_{\rm AF}$ can also describe dimerizations other than
AF (spin-Pierls, charge-density waves).
	That is, there is no explicit spin-dependence in the 
dispersions of the bands in this Appendix.

	The vertex functions important for the present analysis
can be shown in terms of the auxiliary phase defined by
\begin{eqnarray}
\tan \varphi ({\bf k}) &=& 
\frac{ 2\Delta ({\bf k})}{E_c ({\bf k}) - E_{c} ({\bf k} \pm {\bf Q}_{\rm AF}) }.
\label{eqB1}
\end{eqnarray}
%
%
	The   static Raman vertex  and the  current
vertices relevant to both the effective numbers (\ref{eq43})--(\ref{eq44}) and 
the optical conductivity (\ref{eq45}) are given, respectively, by
\begin{eqnarray}
\gamma ^{CC}_{\alpha \alpha} ({\bf k}) 
& =& 
\gamma ^{cc}_{\alpha \alpha} ({\bf k}) \cos^2 \frac{\varphi ({\bf k})}{2}
+ \gamma ^{cc}_{\alpha \alpha} ({\bf k}\pm {\bf Q}_{\rm AF})  
\sin^2 \frac{\varphi ({\bf k})}{2}
\nonumber \\
&&
 -
\frac{m}{e^2}  
\frac{2 |J_{\alpha}^{C\underline{C}} ({\bf k})|^2 }{
 E_{C\underline{C}} ({\bf k}) }, 
 \label{eqB2} 
 \end{eqnarray}
and
\begin{eqnarray}
J^{CC}_{\alpha} ({\bf k}) &=& J^{cc}_{\alpha} ({\bf k}) 
\cos^2 \frac{\varphi ({\bf k})}{2}
+J^{cc}_{\alpha} ({\bf k} \pm {\bf Q}_{\rm AF} ) 
\sin^2 \frac{\varphi ({\bf k})}{2},
\nonumber \\
J^{C\underline{C}}_{\alpha} ({\bf k}) &=& \frac{1}{2}[J^{cc}_{\alpha} ({\bf k}) 
-J^{cc}_{\alpha} ({\bf k} \pm {\bf Q}_{\rm AF} )] 
\sin \varphi ({\bf k}) . 
\label{eqB3}
\end{eqnarray}
%
%
	Here 
$E_{C\underline{C}} ({\bf k}) = E_{C} ({\bf k}) -E_{\underline{C}} ({\bf k})$ 
and 
\begin{eqnarray}
E_{C,\underline{C}} ({\bf k}) &=& 
\frac{1}{2} [E_c ({\bf k}) + E_{c} ({\bf k} \pm {\bf Q}_{\rm AF}  )] 
\label{eqB4} \\  \nonumber
& & 
\pm \sqrt{ \frac{1}{4} [E_c ({\bf k}) - E_{c} ({\bf k}\pm {\bf Q}_{\rm AF} )]^2   
+ \Delta^2({\bf k})}. 
\end{eqnarray}
%
%
	Similarly, the approximate expressions for the total Raman vertices
are given by the expressions (\ref{eq49}).

\section{Longitudinal response theory in multiband models}
%
%

We consider the Hamiltonian (\ref{eq1}) with $H_2' = 0$ and 
$H^{\rm ext}$ given by Eq.~(\ref{eq11}).
 	$H_1'$ includes only the quasi-elastic scattering processes
on the disorder.
	We introduce the retarded electron-hole propagator
${\cal D}^{LL'} ({\bf k}, {\bf k}_+, {\bf k}'_+,{\bf k}',  t)$ 
defined by (hereafter ${\bf q} = q_{\alpha} \hat{e}_{\alpha}$)
\begin{eqnarray}
&&{\cal D}^{LL'} ({\bf k}, {\bf k}_+, {\bf k}'_+,{\bf k}',  t) 
\label{eqC1} \\ \nonumber 
&& \hspace{5mm} = -{\rm i} \Theta(t) \langle \big[ 
 L_{{\bf k} \sigma}^{\dagger} (t) L'_{{\bf k}+{\bf q}  \sigma} (t), 
{L'}_{{\bf k}' +{\bf q} \sigma}^{\dagger}(0) L_{{\bf k}' \sigma} (0)
] \rangle,
\end{eqnarray}
%
%
and the related induced density
\begin{eqnarray}
&& \! \!  \! \! \! \!  \! \! 
\delta n^{LL'} ({\bf k}, {\bf k}_+, \omega)  \equiv \delta 
n^{LL'} ({\bf k})
\label{eqC2} \\ \nonumber
&&
= \sum_{{\bf k}'} \frac{1}{\hbar}
{\cal D}^{LL'} ({\bf k}, {\bf k}_+, {\bf k}'_+,{\bf k}',\omega)
q^{L'L} ({\bf k}'_+,{\bf k}') V^{\rm ext} ({\bf q}, \omega).
\end{eqnarray}
%
%
	The equation of motion for 
${\cal D}^{LL'} ({\bf k}, {\bf k}_+, {\bf k}'_+,{\bf k}', t)$ 
can be set into a form analogous to the Landau equation
\begin{eqnarray}
&&\big[ \hbar \omega  + E_L({\bf k}) 
- E_{L'}({\bf k}_+) \big] \delta n^{LL'} ({\bf k}) 
\nonumber \\ 
&& \hspace{10mm} 
=\big[ f_L({\bf k}) - f_{L'}({\bf k}_+) \big] q^{L'L} ({\bf k}_+,{\bf k}) 
V^{\rm ext} ({\bf q}, \omega)
\nonumber \\ 
&& \hspace{15mm} 
- {\rm i} \hbar {\rm Im} \{\Sigma_{\alpha}^{LL'} ({\bf k},\omega) \}
\delta \tilde{n}^{LL'} ({\bf k}),
\label{eqC3} 
\end{eqnarray}
%
%
where $\delta \tilde{n}^{LL'} ({\bf k})$ is the contribution to
$\delta n^{LL'} ({\bf k})$ which is proportional to 
$J^{L'L}_{\alpha} ({\bf k})$ and 
\begin{eqnarray}
&& \hbar \Sigma_{\alpha}^{LL'}({\bf k},\omega) \approx 
-\sum_{{\bf q}'}\big|V_1 ({\bf q}') \big|^2  \frac{1}{\hbar}
\big[  {\cal D}_0^{LL'} ({\bf k}, {\bf k}_++{\bf q}', \omega) 
\nonumber \\ 
&& \hspace{5mm} +  
 {\cal D}_0^{LL'} ({\bf k}+{\bf q}', {\bf k}_+, \omega) \big] 
\bigg( 1 - 
\frac{J^{L'L}_{\alpha} ({\bf k}+{\bf q}')}{ J^{L'L}_{\alpha} ({\bf k})} \bigg)
\label{eqC4} 
\end{eqnarray}
%
%
is the  electron-hole self-energy for the case 
$V_1^{LL}({\bf q}') \approx V_1 ({\bf q}')$, and
\begin{eqnarray}
\frac{1}{\hbar} {\cal D}_0^{LL'}({\bf k},{\bf k}',\omega ) &=&
\frac{1}{\hbar \omega + E_L({\bf k}) - E_{L'}({\bf k}') + {\rm i} \hbar \eta}.
\nonumber \\ 
\label{eqC5} 
\end{eqnarray}
%
%
	In  expression (\ref{eqC3}) the fact
that the real part of the electron-hole self-energy is negligible 
for the quasi-elastic scattering on disorder is taken into account.

The total induced density $\delta n^{LL'} ({\bf k})$ 
consists of the induced charge and 
current densities [denoted  by $\delta n^{LL'}_0 ({\bf k})$
and $\delta n^{LL'}_1 ({\bf k})$ \cite{Pines}],
satisfying the (intraband) continuity equation
$\hbar \omega \delta n^{LL}_0 ({\bf k}) + E_{LL}({\bf k},{\bf k}_+)
\delta n^{LL}_1 ({\bf k}) = 0$. 
	The solution of the  Landau equation (\ref{eqC3}),
together with the definition for the total optical conductivity
\begin{equation}
j_{\alpha}^{\rm ind} (\omega) = \frac{1}{v} \sum_{LL' {\bf k} \sigma}
J_{\alpha}^{LL'} ({\bf k}) \delta n^{LL'}_1 ({\bf k})
= \sigma_{\alpha \alpha} (\omega) E^{\rm ext}_{\alpha} (\omega) 
\label{eqC6} 
\end{equation}
%
%
and with the relation
\begin{equation}
q^{L'L} ({\bf k}_+,{\bf k}) V^{\rm ext} ({\bf q}, \omega)
\approx  
\frac{\hbar J_{\alpha}^{L'L} ({\bf k})}{
E_{L'L}({\bf k}_+,{\bf k})} {\rm i} E^{\rm ext}_{\alpha} (\omega),
\label{eqC7} 
\end{equation}
%
%
[corresponding to Eq.~(\ref{eq12}) combined with  the relation
$q_{\alpha} V^{\rm ext} ({\bf q},\omega) = 
{\rm i} E^{\rm ext}_{\alpha} (\omega)$]  gives 
\begin{eqnarray}
&& \! \!  \! \! \! \!  \! \! 
\sigma_{\alpha \alpha} (\omega) = \frac{\rm i}{\omega}
\frac{1}{v} \sum_{LL' {\bf k} \sigma}
\bigg( \frac{\hbar \omega }{E_{L'L}({\bf k}_+,{\bf k}) } \bigg)^{n_{LL'}}
\big| J_{\alpha}^{LL'} ({\bf k}) \big|^2
\nonumber \\
&& \! \!  \! \! \! \!  \! \! \! \!  \! \!   
\times \frac{ f_L({\bf k}) - f_{L'}({\bf k}_+)}{
\hbar \omega + {\rm i}\hbar \Gamma_{\alpha}^{LL'}({\bf k},\omega) 
+ E_{LL'}({\bf k},{\bf k}) 
- \displaystyle \frac{E_{L'L'}^2({\bf k},{\bf k}_+)}{\hbar \omega}}.
\label{eqC8}
\end{eqnarray}
%
%
	Here $n_{LL} = 1$ in the intraband channel, $n_{L\underline{L}} = 2$ 
in the interband channel,
$\Gamma_{\alpha}^{LL'}({\bf k},\omega) = 
{\rm Im}\{\Sigma_{\alpha}^{LL'}({\bf k},\omega)\}$ and  
$E_{LL'}({\bf k},{\bf k}') = E_{L}({\bf k})-E_{L'}({\bf k}')$.
	The related long-wavelength susceptibility and  the 
dielectric function become
\begin{eqnarray}
e^2\chi_{1,1} ({\bf q}, \omega) &=& 
-\sum_{\alpha} \frac{{\rm i} q_{\alpha}^2}{\omega}\sigma_{\alpha \alpha} (\omega),
\nonumber \\
\varepsilon({\bf q}, \omega) &=& 
1 + \frac{4 \pi {\rm i}}{\omega q^2} \sum_{\alpha} 
q_{\alpha}^2\sigma_{\alpha \alpha}(\omega),
\label{eqC9} 
\end{eqnarray}
%
%
with ${\bf q} = \sum_{\alpha} q_{\alpha} \hat{e}_{\alpha}$.
	The expressions (\ref{eqC8})--(\ref{eqC9}) are the generalization
of the well-known single-band Landau response functions \cite{Pines}. 
	Obviously, to obtain Eqs.~(\ref{eq22})--(\ref{eq23})
of the main text we have to include the contributions beyond 
the three-band  model, as well,
by adding $\varepsilon_{\infty}({\bf q}, \omega) - 1$
to the above expression for 
$\varepsilon({\bf q}, \omega)$.


\end{document}